\documentclass[english]{IEEEtran}
\pdfoutput=1
\usepackage[T1]{fontenc}
\usepackage[latin9]{inputenc}
\usepackage{color}
\usepackage{babel}

\usepackage{enumitem}
\usepackage{amsthm}
\usepackage{amsmath}
\usepackage{amssymb}
\usepackage{xargs}[2008/03/08]
\usepackage[unicode=true,
 bookmarks=true,bookmarksnumbered=false,bookmarksopen=false,
 breaklinks=true,pdfborder={0 0 0},backref=false,colorlinks=true]
 {hyperref}
\hypersetup{pdftitle={Regular Combinators for String Transformations},
 pdfauthor={Rajeev Alur, Adam Freilich, Mukund Raghothaman},
 pdfsubject={Transducers},
 pdfkeywords={String transducers, MSO, Expression languages}}

\makeatletter
\numberwithin{equation}{section}
\numberwithin{figure}{section}
\theoremstyle{plain}
\newtheorem{thm}{\protect\theoremname}
\theoremstyle{definition}
\newtheorem{example}[thm]{\protect\examplename}
\theoremstyle{plain}
\newtheorem{prop}[thm]{\protect\propositionname}
\theoremstyle{definition}
\newtheorem{defn}[thm]{\protect\definitionname}
\theoremstyle{plain}
\newtheorem{lem}[thm]{\protect\lemmaname}
\theoremstyle{remark}
\newtheorem{claim}[thm]{\protect\claimname}
 \newlist{casenv}{enumerate}{4}
 \setlist[casenv]{leftmargin=*,align=left,widest={iiii}}
 \setlist[casenv,1]{label={{\itshape\ \casename} \arabic*.},ref=\arabic*}
 \setlist[casenv,2]{label={{\itshape\ \casename} \roman*.},ref=\roman*}
 \setlist[casenv,3]{label={{\itshape\ \casename\ \alph*.}},ref=\alph*}
 \setlist[casenv,4]{label={{\itshape\ \casename} \arabic*.},ref=\arabic*}

\usepackage{amssymb}
\usepackage{bussproofs}
\usepackage{centernot}
\usepackage{datetime}
\usepackage{filecontents}
\usepackage{fullpage}
\usepackage{microtype}
\usepackage{tikz}
\usepackage{pgfplots}
\usepackage{stmaryrd}

\usetikzlibrary{arrows}
\usetikzlibrary{automata}
\usetikzlibrary{backgrounds}
\usetikzlibrary{fit}
\usetikzlibrary{positioning}

\@ifundefined{showcaptionsetup}{}{%
 \PassOptionsToPackage{caption=false}{subfig}}
\usepackage{subfig}
\makeatother

 \providecommand{\casename}{Case}
\providecommand{\claimname}{Claim}
\providecommand{\definitionname}{Definition}
\providecommand{\examplename}{Example}
\providecommand{\lemmaname}{Lemma}
\providecommand{\propositionname}{Proposition}
\providecommand{\theoremname}{Theorem}

\begin{document}

\title{Regular Combinators for String Transformations}

\author{\IEEEauthorblockN{Rajeev Alur, Adam Freilich, Mukund Raghothaman}\\
\IEEEauthorblockA{University of Pennsylvania}}

\IEEEspecialpapernotice{This is the full version of the paper, and includes proofs omitted
from the short version.}

\maketitle

\global\long\def\autobox#1{#1}

\global\long\def\opname#1{\autobox{\operatorname{#1}}}

\global\long\def\dontcare{\_}

\global\long\def\bbracket#1{\left\llbracket #1\right\rrbracket }

\global\long\def\cnot#1{\centernot#1}

\global\long\def\p#1{(#1)}

\global\long\def\proofsep{\mathbin{\vdash}}

\global\long\def\fCenter{\proofsep}

\global\long\def\binaryprimitive#1#2{\BinaryInf$#1\fCenter#2$}

\global\long\def\axiomprimitive#1#2{\Axiom$#1\fCenter#2$}

\global\long\def\unaryprimitive#1#2{\UnaryInf$#1\fCenter#2$}

\global\long\def\trinaryprimitive#1#2{\TrinaryInf$#1\fCenter#2$}

\global\long\def\bussproof#1{#1\DisplayProof}

\global\long\def\binaryinfc#1#2#3#4{#1#2\RightLabel{\ensuremath{{\scriptstyle \textrm{#4}}}}\BinaryInfC{\ensuremath{#3}}}

\global\long\def\trinaryinfc#1#2#3#4#5{#1#2#3\RightLabel{\ensuremath{{\scriptstyle \textrm{#5}}}}\TrinaryInfC{\ensuremath{#4}}}

\global\long\def\unaryinfc#1#2#3{#1\RightLabel{\ensuremath{{\scriptstyle \textrm{#3}}}}\UnaryInfC{\ensuremath{#2}}}

\global\long\def\axiomc#1{\AxiomC{\ensuremath{#1}}}

\global\long\def\binaryinf#1#2#3#4#5{#1#2\RightLabel{\ensuremath{{\scriptstyle \textrm{#5}}}}\binaryprimitive{#3}{#4}}

\global\long\def\trinaryinf#1#2#3#4#5#6{#1#2#3\RightLabel{\ensuremath{{\scriptstyle \textrm{#6}}}}\trinaryprimitive{#4}{#5}}

\global\long\def\unaryinf#1#2#3#4{#1\RightLabel{\ensuremath{{\scriptstyle \textrm{#4}}}}\unaryprimitive{#2}{#3}}

\global\long\def\axiom#1#2{\axiomprimitive{#1}{#2}}

\global\long\def\axrulesp#1#2#3#4{\bussproof{\unaryinf{\axiomc{\vphantom{#4}}}{#1}{#2}{#3}}}

\global\long\def\axrule#1#2#3{\axrulesp{#1}{#2}{#3}{\proofsep\Gamma,\Delta}}

\global\long\def\unrule#1#2#3#4#5{\bussproof{\unaryinf{\axiom{#1}{#2}}{#3}{#4}{#5}}}

\global\long\def\unrulec#1#2#3{\bussproof{\unaryinfc{\axiomc{#1}}{#2}{#3}}}

\global\long\def\binrule#1#2#3#4#5#6#7{\bussproof{\binaryinf{\axiom{#1}{#2}}{\axiom{#3}{#4}}{#5}{#6}{#7}}}

\global\long\def\binrulec#1#2#3#4{\bussproof{\binaryinfc{\axiomc{#1}}{\axiomc{#2}}{#3}{#4}}}

\global\long\def\roset#1{\left\{  #1\right\}  }

\global\long\def\rosetbr#1{\{#1\}}

\global\long\def\ruset#1#2{\roset{#1\;\middle\vert\;#2}}

\global\long\def\rusetbr#1#2{\{#1\;\vert\;#2\}}

\global\long\def\union#1#2{#1\cup#2}

\global\long\def\bigunion#1#2{\bigcup_{#1}#2}

\global\long\def\intersection#1#2{#1\cap#2}

\global\long\def\bigintersection#1#2{\bigcap_{#1}#2}

\global\long\def\bigland#1#2{\bigwedge_{#1}#2}

\global\long\def\powerset#1{2^{#1}}

\global\long\def\cart#1#2{#1\times#2}

\global\long\def\tuple#1{\left(#1\right)}

\global\long\def\brtuple#1{(#1)}

\global\long\def\quoset#1#2{\left.#1\middle/#2\right.}

\global\long\def\equivclass#1{\left[#1\right]}

\global\long\def\refltransclosure#1{#1^{*}}

\newcommandx\funcapphidden[3][usedefault, addprefix=\global, 1=]{#2#1#3}

\global\long\def\funcapplambda#1#2{\funcapphidden[\,]{#1}{#2}}

\global\long\def\funcapptrad#1#2{\funcapphidden{#1}{\tuple{#2}}}

\global\long\def\fabr#1#2{\funcapphidden{#1}{\brtuple{#2}}}

\global\long\def\funccomp#1#2{#1\circ#2}

\global\long\def\arrow#1#2{#1\to#2}

\global\long\def\func#1#2#3{#1:\arrow{#2}{#3}}

\global\long\def\N{\mathbb{N}}

\global\long\def\Z{\mathbb{Z}}

\global\long\def\Q{\mathbb{Q}}

\global\long\def\R{\mathbb{R}}

\global\long\def\D{\mathbb{D}}

\global\long\def\Zmod#1{\quoset{\Z}{#1\Z}}

\global\long\def\vector#1{\mathbf{#1}}

\global\long\def\dotprod#1#2{#1\cdot#2}

\global\long\def\bool{{\tt bool}}

\global\long\def\true{{\tt true}}

\global\long\def\false{{\tt false}}

\global\long\def\inlinemod#1#2{#1\bmod{#2}}

\global\long\def\parenmod#1#2{#1\pmod{#2}}

\global\long\def\strempty{\epsilon}

\newcommandx\strcat[3][usedefault, addprefix=\global, 1=]{#2#1#3}

\global\long\def\strrev#1{#1^{\autobox{\mathit{rev}}}}

\global\long\def\strlen#1{\left|#1\right|}

\global\long\def\strlenp#1#2{\strlen{#1}_{#2}}

\global\long\def\prefix#1#2{\funcapptrad{\texttt{\ensuremath{\opname{pre}}}}{#1,#2}}

\global\long\def\suffix#1#2{\funcapptrad{\texttt{suf}}{#1,#2}}

\global\long\def\strproj#1#2{\funcapptrad{\pi_{#2}}{#1}}

\global\long\def\regexor#1#2{#1+#2}

\newcommandx\regexconcat[3][usedefault, addprefix=\global, 1=]{\strcat[#1]{#2}{#3}}

\global\long\def\kstar#1{#1^{*}}

\global\long\def\kplus#1{#1^{+}}

\global\long\def\bigoh#1{\funcapptrad O{#1}}

\global\long\def\littleoh#1{\funcapptrad o{#1}}

\global\long\def\bigomega#1{\funcapptrad{\Omega}{#1}}

\global\long\def\littleomega#1{\funcapptrad{\omega}{#1}}

\global\long\def\bigtheta#1{\funcapptrad{\Theta}{#1}}

\global\long\def\logspace{\textsc{logspace}}

\global\long\def\nlogspace{\textsc{nlogspace}}

\global\long\def\np{\textsc{np}}

\global\long\def\conp{\textsc{co}\np}

\global\long\def\pspace{\textsc{pspace}}

\global\long\def\npspace{\textsc{npspace}}

\global\long\def\exptime{\textsc{exptime}}

\global\long\def\complete{\mbox{-complete}}

\global\long\def\hard{\mbox{-hard}}

\global\long\def\npc{\np\complete}

\global\long\def\nph{\np\hard}

\global\long\def\conpc{\conp\complete}

\global\long\def\pspacec{\pspace\complete}

\global\long\def\pspaceh{\pspace\hard}

\global\long\def\exptimec{\exptime\complete}

\global\long\def\exptimeh{\exptime\hard}

\global\long\def\crasimp{\operatorname{\mbox{CRA}}}

\global\long\def\cra#1{\funcapptrad{\crasimp}{#1}}

\global\long\def\ccrasimp{\operatorname{\mbox{CCRA}}}

\global\long\def\ccra#1{\funcapptrad{\ccrasimp}{#1}}

\global\long\def\sst{\operatorname{\mbox{SST}}}

\global\long\def\acra{\operatorname{\mbox{ACRA}}}

\global\long\def\dacra#1{\funcapptrad{\acra}{#1}}

\global\long\def\wa{\operatorname{\mbox{WA}}}

\newcommandx\oracleeval[1][usedefault, addprefix=\global, 1=\cdot]{\operatorname{\textsc{EVAL}}\left(#1\right)}

\newcommandx\oracleequiv[1][usedefault, addprefix=\global, 1=\cdot]{\operatorname{\textsc{EQUIV}}\left(#1\right)}

\global\long\def\interp#1{\bbracket{#1}}

\global\long\def\domain#1{\funcapptrad{\autobox{\mbox{Dom}}}{#1}}

\newcommand{\branchshortorfull}[1]{}

\branchshortorfull{Neither branch enabled.}

\global\long\def\choiceop{\vartriangleright}

\global\long\def\repsumop{+}

\global\long\def\splitsumop{\oplus}

\global\long\def\lsplitsumop{\overleftarrow{\splitsumop}}

\global\long\def\itersumop{\sum}

\global\long\def\litersumop{\overleftarrow{\itersumop}}

\global\long\def\compop{\circ}

\global\long\def\botfn{\bot}

\global\long\def\const#1#2{\left.#1\middle/#2\right.}

\global\long\def\choice#1#2{#1\choiceop#2}

\global\long\def\repsum#1#2{#1\repsumop#2}

\global\long\def\splitsum#1#2{#1\splitsumop#2}

\global\long\def\lsplitsum#1#2{#1\lsplitsumop#2}

\global\long\def\itersum#1{\itersumop#1}

\global\long\def\litersum#1{\litersumop#1}

\global\long\def\funrev#1{\strrev{#1}}

\global\long\def\comp#1#2{#1\compop#2}

\global\long\def\restrict#1#2{\intersection{#1}{#2}}

\global\long\def\lshift#1#2{#1\ll#2}

\global\long\def\rshift#1#2{#1\gg#2}

\global\long\def\trivconst#1{#1}

\global\long\def\indicatorfn#1{{\bf 1}_{#1}}

\global\long\def\idfn{\autobox{\mathit{id}}}

\global\long\def\repeatfn{\autobox{\mathit{copy}}}

\global\long\def\coffee{\autobox{\mathit{coffee}}}

\global\long\def\swapfn{\autobox{\mathit{swap}}}

\global\long\def\stripfn{\autobox{\mathit{strip}}}

\global\long\def\shuffleexfn{\autobox{\mathit{shuffle}}}

\global\long\def\chainsum#1#2{\itersum{\tuple{#1,#2}}}

\global\long\def\lchainsum#1#2{\litersum{\tuple{#1,#2}}}

\global\long\def\subcalc#1{\funcapptrad{\mathbb{F}}{#1}}

\begin{abstract}
We focus on (partial) functions that map input strings to a monoid
such as the set of integers with addition and the set of output strings
with concatenation. The notion of regularity for such functions has
been defined using two-way finite-state transducers, (one-way) cost
register automata, and MSO-definable graph transformations. In this
paper, we give an algebraic and machine-independent characterization
of this class analogous to the definition of regular languages by
regular expressions. When the monoid is commutative, we prove that
every regular function can be constructed from constant functions
using the combinators of choice, split sum, and iterated sum, that
are analogs of union, concatenation, and Kleene-{*}, respectively,
but enforce unique (or unambiguous) parsing. Our main result is for
the general case of non-commutative monoids, which is of particular
interest for capturing regular string-to-string transformations for
document processing. We prove that the following additional combinators
suffice for constructing all regular functions: (1) the left-additive
versions of split sum and iterated sum, which allow transformations
such as string reversal; (2) sum of functions, which allows transformations
such as copying of strings; and (3) function composition, or alternatively,
a new concept of chained sum, which allows output values from adjacent
blocks to mix.
\end{abstract}

\section{Introduction \label{sec:Intro}}

To study string transformations, given the success of finite-state
automata and the associated theory of regular languages, a natural
starting point is the model of finite-state transducers. A finite-state
transducer emits output symbols at every step, and given an input
string, the corresponding output string is the concatenation of all
the output symbols emitted by the machine during its execution. Such
transducers have been studied since the 1960s, and it has been known
that the transducers have very different properties compared to the
acceptors: \emph{two-way} transducers are strictly more expressive
than their one-way counter-parts, and the post-image of a regular
language under a two-way transducer need not be a regular language
\cite{AHU69}. For the class of transformations computed by two-way
transducers, \cite{CJ77} establishes closure under composition, \cite{Gu80}
proves decidability of functional equivalence, and \cite{EH01} shows
that their expressiveness coincides with MSO-definable string-to-string
transformations of \cite{Cou92}. As a result, \cite{EH01} justifiably
dubbed this class as \emph{regular} string transformations. Recently,
an alternative characterization using one-way machines was found for
this class: \emph{streaming string transducers} \cite{AC10} (and
their more general and abstract counterpart of \emph{cost register
automata} \cite{CRA-LICS}) process the input string in a single left-to-right
pass, but use multiple write-only registers to store partially computed
output chunks that are updated and combined to compute the final answer.

There has been a resurgent interest in such transducers in the formal
methods community with applications to learning of string transformations
from examples \cite{Gul11}, sanitization of web addresses \cite{VHLMB12},
and algorithmic verification of list-processing programs \cite{SST-POPL}.
In the context of these applications, we wish to focus on regular
transformations, rather than the subclass of classical one-way transducers,
since the gap includes many natural transformations such as string
reversal and swapping of substrings, and since one-way transducers
are not closed under basic operations such as choice.

For our formal study, we focus on \emph{cost functions}, that is,
(partial) functions that map strings over a finite alphabet to values
from a monoid $\tuple{\D,+,0}$. While the set of output strings with
concatenation is a typical example of such a monoid, cost functions
can also associate numerical values (or rewards) with sequences of
events, with possible application to \emph{quantitative} analysis
of systems \cite{CDH-10} (it is worth pointing out that the notion
of regular cost functions proposed by Colcombet is quite distinct
from ours \cite{Col09}). An example of such a numerical domain is
the set of integers with addition. In the case of a \emph{commutative}
monoid, regular functions have a simpler structure, and correspond
to \emph{unambiguous weighted automata} (note that weighted automata
are generally defined over a semiring, and are very extensively studied---see
\cite{DKV09} for a survey, but with no results directly relevant
to our purpose). As another interesting example of a numerical monoid, each value is
a cost-discount pair, and the (non-commutative) addition is the discounted
sum operation. The traditional use of discounting in systems theory
allows only discounting of \emph{future} events, and corresponds to
cost functions computed by classical one-way transducers, while regular
functions allow more general forms of discounting (for instance, discounting
of both past and future events).

A classical result in automata theory characterizes regular languages
using \emph{regular expressions}: regular languages are exactly the
sets that can be inductively generated from base languages (empty
set, empty string, and alphabet symbols) using the operations of union,
concatenation, and Kleene-{*}. Regular expressions provide a robust
foundation for specifying regular patterns in a \emph{declarative}
manner, and are widely used in practical applications. The goal of
this paper is to identify the appropriate base functions and combinators
over cost functions for an analogous algebraic and machine-independent
characterization of regularity.

We begin our study by defining base functions and combinators that
are the analogs of the classical operations used in regular expressions.
The base function $\const Ld$ maps strings $\sigma$ in the base
language $L$ to the constant value $d$, and is undefined when $\sigma\notin L$.
Given cost functions $f$ and $g$, the \emph{conditional choice}
combinator $\choice fg$ maps an input string $\sigma$ to $\funcapptrad f{\sigma}$,
if this value is defined, and to $\funcapptrad g{\sigma}$ otherwise;
the \emph{split sum} combinator $\splitsum fg$ maps an input string
$\sigma$ to $\funcapptrad f{\sigma_{1}}+\funcapptrad f{\sigma_{2}}$
if the string $\sigma$ can be split \emph{uniquely} into two parts
$\sigma_{1}$ and $\sigma_{2}$ such that both $\funcapptrad f{\sigma_{1}}$
and $\funcapptrad g{\sigma_{2}}$ are defined, and is undefined otherwise;
and the \emph{iterated sum} $\itersum f$ is defined so that if the
input string $\sigma$ can be split uniquely such that $\sigma=\sigma_{1}\sigma_{2}\ldots\sigma_{k}$
and each $\funcapptrad f{\sigma_{i}}$ is defined, then $\funcapptrad{\itersum f}{\sigma}$
is $\funcapptrad f{\sigma_{1}}+\funcapptrad f{\sigma_{2}}+\cdots+\funcapptrad f{\sigma_{k}}$,
and is undefined otherwise. The combinators conditional choice, split
sum, and iterated sum are the natural analogs of the operations of
union, concatenation, and Kleene-{*} over languages, respectively.
The uniqueness restrictions ensure that the input string is parsed
in an unambiguous manner while computing its cost, and thus, the result
of combining two (partial) functions remains a (partial) \emph{function}.

Our first result is that when the operation $+$ is commutative, regular
functions are exactly the functions that can be inductively generated
from base functions using the combinators of conditional choice, split
sum, and iterated sum. The proof is fairly straightforward, and builds
on the known properties of cost register automata, their connection
to unambiguous weighted automata in the case of commutative monoids,
and the classical translation from automata to regular expressions.

When the operation $+$ is not commutative, which is the case when
the output values are strings themselves and addition corresponds
to string concatenation, we need additional combinators to capture
regularity. First, in the non-commutative case, it is natural to introduce
symmetric \emph{left-additive} versions of split sum and iterated
sum. Given cost functions $f$ and $g$, the \emph{left-split sum}
$\lsplitsum fg$ maps an input string $\sigma$ to $\funcapptrad g{\sigma_{2}}+\funcapptrad f{\sigma_{1}}$
if the string $\sigma$ can be split uniquely into two parts $\sigma_{1}$
and $\sigma_{2}$ such that both $\funcapptrad f{\sigma_{1}}$ and
$\funcapptrad g{\sigma_{2}}$ are defined. The \emph{left-iterated
sum} is defined analogously, and in particular, the transformation
that maps an input string to its \emph{reverse} is simply the left-iterated
sum of the function that maps each symbol to itself. It is easy to
show that regular functions are closed under these left-additive combinators.

The \emph{sum} $\repsum fg$ of two functions $f$ and $g$ maps a
string $\sigma$ to $\funcapptrad f{\sigma}+\funcapptrad g{\sigma}$.
Though the sum combinator is not necessary for completeness in the
commutative case, it is natural for cost functions. For example, the
\emph{string copy} function that maps an input string $\sigma$ to
the output $\sigma\sigma$ is simply the sum of the identity function
over strings with itself. It is already known that regular functions
are closed under sum \cite{EH01,CRA-LICS}.

To motivate our final combinator, consider the string-transformation
$\shuffleexfn$ that maps a string of the form $a^{m_{1}}ba^{m_{2}}b\ldots a^{m_{k}}b$
to $a^{m_{2}}b^{m_{1}}a^{m_{3}}b^{m_{2}}\ldots a^{m_{k}}b^{m_{k-1}}$.
This function is definable using cost register automata, but we conjecture
that it cannot be constructed using the combinators discussed so far.
We introduce a new form of iterated sum: given a language $L$ and
a cost function $f$, if the input string $\sigma$ can be split uniquely
so that $\sigma=\sigma_{1}\sigma_{2}\ldots\sigma_{k}$ with each $\sigma_{i}\in L$,
then the \emph{chained sum} $\chainsum fL$ of $\sigma$ is $\funcapptrad f{\sigma_{1}\sigma_{2}}+\funcapptrad f{\sigma_{2}\sigma_{3}}+\cdots+\funcapptrad f{\sigma_{k-1}\sigma_{k}}$.
In other words, the input is (uniquely) divided into substrings belonging
to the language $L$, but instead of summing the values of $f$ on
each of these substrings, we sum the values of $f$ applied to blocks
of adjacent substrings in a chained fashion. The string-transformation
$\shuffleexfn$ now is simply chained sum where $L$ equals the regular
language $\kstar ab$, and $f$ maps $a^{i}ba^{j}b$ to $a^{j}b^{i}$
(such a function $f$ can be constructed using iterated sum and left-split
sum). It turns out that this new combinator can also be defined if
we allow \emph{function composition}: if $f$ is a function that maps
strings to strings and $g$ is a cost function, then the composed
function $\comp gf$ maps an input string $\sigma$ to $\funcapptrad g{\funcapptrad f{\sigma}}$.
Such rewriting is a natural operation, and regular functions are closed
under composition \cite{CJ77}.

The main technical result of the paper is that every regular function
can be inductively generated from base functions using the combinators
of conditional choice, sum, split sum, either chained sum or function
composition, and their left additive versions. The proof in section
\ref{sec:Noncomm} constructs the desired expressions corresponding
to executions of cost register automata. Such automata have multiple
registers, and at each step the registers are updated using \emph{copyless}
(or single-use) assignments. Register values can flow into one another
in a complex manner, and the proof relies on understanding the structure
of compositions of \emph{shapes} that capture these value-flows. The
proof provides insights into the power of the chained sum operation,
and also offers an alternative justification for the copyless restriction
for register updates in the machine-based characterization of regular
functions.

\section{Function Combinators \label{sec:Combinators}}

Let $\Sigma$ be a finite alphabet, and $\tuple{\D,+,0}$ be a monoid.
Two natural monoids of interest are those of the integers $\tuple{\Z,+,0}$
under addition, and of strings $\tuple{\kstar{\Gamma},\cdot,\strempty}$
over some output alphabet $\Gamma$ under concatenation. By convention,
we treat $\bot$ as the undefined value, and express partial functions
$\func fAB$ as total functions $\func fA{B_{\bot}}$, where $B_{\bot}=\union B{\roset{\bot}}$.
We extend the semantics of the monoid $\D$ to $\D_{\bot}$ by defining
$d+\bot=\bot+d=\bot$, for all $d\in\D$. A \emph{cost function} is
a function $\arrow{\kstar{\Sigma}}{\D_{\bot}}$.

\subsection{Base functions \label{sub:Combinators:Base}}

For each language $L\subseteq\kstar{\Sigma}$ and $d\in\D$, we define
the \emph{constant function $\func{\const Ld}{\kstar{\Sigma}}{\D_{\bot}}$}
as 
\begin{alignat*}{1}
\funcapptrad{\const Ld}{\sigma} & =\begin{cases}
d & \mbox{if }\sigma\in L,\mbox{ and}\\
\bot & \mbox{otherwise}.
\end{cases}
\end{alignat*}
The \emph{everywhere-undefined function $\func{\botfn}{\kstar{\Sigma}}{\D_{\bot}}$}
is defined as $\funcapptrad{\botfn}{\sigma}=\bot$. $\botfn$ can
also be defined as the constant function $\const{\emptyset}0$.
\begin{example}
\label{ex:Combinators:ConstPoint} Let $\Sigma=\roset{a,b}$ in the
following examples. Then, the constant function $\func{\const aa}{\kstar{\Sigma}}{\kstar{\Sigma}}$maps
$a$ to itself, and is undefined on all other strings. We will often
be interested in functions of the form $\const aa$: when the intent
is clear, we will use the shorthand $\trivconst a$.
\end{example}
By \emph{base functions}, we refer to the class of functions $\const Ld$,
where $L$ is a regular language.

\subsection{Conditional choice and sum operators \label{sub:Combinators:CondSum}}

Let $\func{f,g}{\kstar{\Sigma}}{\D}_{\bot}$ be two functions. We
then define the \emph{conditional choice $\choice fg$} as 
\begin{alignat*}{1}
\funcapptrad{\choice fg}{\sigma} & =\begin{cases}
\funcapptrad f{\sigma} & \mbox{if }\funcapptrad f{\sigma}\neq\bot,\mbox{ and}\\
\funcapptrad g{\sigma} & \mbox{otherwise}.
\end{cases}
\end{alignat*}

\begin{example}
\label{ex:Combinators:CondSum} The indicator function $\func{\indicatorfn L}{\kstar{\Sigma}}{\Z}$
is defined as $\funcapptrad{\indicatorfn L}{\sigma}=1$ if $\sigma\in L$
and $\funcapptrad{\indicatorfn L}{\sigma}=0$ otherwise. This function
can be expressed using the conditional choice operator as $\choice{\const L1}{\const{\kstar{\Sigma}}0}$.
\end{example}
The \emph{sum $\repsum fg$} is defined as $\funcapptrad{\repsum fg}{\sigma}=\funcapptrad f{\sigma}+\funcapptrad g{\sigma}$.
If there exist unique strings $\sigma_{1}$ and $\sigma_{2}$ such
that $\sigma=\sigma_{1}\sigma_{2}$, and $\funcapptrad f{\sigma_{1}}$
and $\funcapptrad g{\sigma_{2}}$ are both defined, then the \emph{split
sum $\funcapptrad{\splitsum fg}{\sigma}=\funcapptrad f{\sigma_{1}}+\funcapptrad g{\sigma_{2}}$}.
Otherwise, $\funcapptrad{\splitsum fg}{\sigma}=\bot$. Over non-commutative
monoids, this may be different from the\emph{ left-split sum $\lsplitsum fg$}:
if there exist unique strings $\sigma_{1}$ and $\sigma_{2}$, such
that $\sigma=\sigma_{1}\sigma_{2}$, and $\funcapptrad f{\sigma_{1}}$
and $\funcapptrad g{\sigma_{2}}$ are both defined, then $\funcapptrad{\lsplitsum fg}{\sigma}=\funcapptrad g{\sigma_{2}}+\funcapptrad f{\sigma_{1}}$.
Otherwise, $\funcapptrad{\lsplitsum fg}{\sigma}=\bot$.

Observe that $\choiceop$ is the analogue of union in regular expressions,
with the important difference being that $\choiceop$ is non-commutative.
Similarly, $\splitsumop$ is similar to the concatenation operator
of traditional regular expressions.

\subsection{Iteration \label{sub:Combinators:Iteration}}

The \emph{iterated sum $\itersum f$} of a cost function is defined
as follows. If there exist unique strings $\sigma_{1}$, $\sigma_{2}$,
\ldots{}, $\sigma_{k}$ such that $\sigma=\sigma_{1}\sigma_{2}\ldots\sigma_{k}$
and $\funcapptrad f{\sigma_{i}}$ is defined for each $\sigma_{i}$,
then $\funcapptrad{\itersum f}{\sigma}=\funcapptrad f{\sigma_{1}}+\funcapptrad f{\sigma_{2}}+\cdots+\funcapptrad f{\sigma_{k}}$.
Otherwise, $\funcapptrad{\itersum f}{\sigma}=\bot$. The \emph{left-iterated
sum $\litersum f$} is defined similarly: if there exist unique strings
$\sigma_{1}$, $\sigma_{2}$, \ldots{}, $\sigma_{k}$ such that $\sigma=\sigma_{1}\sigma_{2}\ldots\sigma_{k}$
and $\funcapptrad f{\sigma_{i}}$ is defined for each $\sigma_{i}$,
then $\funcapptrad{\litersum f}{\sigma}=\funcapptrad f{\sigma_{k}}+\funcapptrad f{\sigma_{k-1}}+\cdots+\funcapptrad f{\sigma_{1}}$.
Otherwise, $\funcapptrad{\litersum f}{\sigma}=\bot$. The \emph{reverse
combinator $\funrev f$} is defined as $\funcapptrad{\funrev f}{\sigma}=\funcapptrad f{\strrev{\sigma}}$.
Observe that the left-iterated sum and reverse combinators are interesting
in the case of non-commutative monoids, such as string concatenation.
\begin{example}
\label{ex:Combinators:Iteration:SimpleNumbers} The function $\func{\strlenp{\cdot}a}{\kstar{\Sigma}}{\Z}$
counts the number of $a$-s in the input string. This is represented
by the function expression $\itersum{\left(\choice{\const a1}{\const b0}\right)}$.
The identity function $\func{\idfn}{\kstar{\Sigma}}{\kstar{\Sigma}}$
is given by the function expression $\itersum{\left(\choice{\trivconst a}{\trivconst b}\right)}$.
The function $\repeatfn$ which maps an input $\sigma$ to $\strcat{\sigma}{\sigma}$
is then given by the expression $\repsum{\idfn}{\idfn}$. On the other
hand, the expression $\litersum{\left(\choice{\trivconst a}{\trivconst b}\right)}$
is the function which reverses its input: $\funcapptrad{\litersum{\left(\choice{\trivconst a}{\trivconst b}\right)}}{\sigma}=\strrev{\sigma}$
for all $\sigma$. This is also equivalent to the expression $\funrev{\idfn}$.
\end{example}

\begin{example}
\label{ex:Combinators:Iteration:Coffee} Consider the situation of
a customer who frequents a coffee shop. Every cup of coffee he purchases
costs $\$2$, but if he fills out a survey, then all cups of coffee
purchased that month cost only $\$1$ (including cups already purchased).
Here $\Sigma=\roset{C,S,\#}$ denoting respectively the purchase of
a cup of coffee, completion of the survey, and the passage of a calendar
month. Then, the function expression $m=\choice{\left(\itersum{\const C2}\right)}{\left(\splitsum{\left(\itersum{\const C1}\right)}{\splitsum{\const S0}{\itersum{\left(\choice{\const C1}{\const S0}\right)}}}\right)}$
maps the purchases of a month to the customer's debt. The first sub-expression
-- $\itersum{\const C2}$ -- computes the amount provided no survey
is filled out and the second sub-expression -- $\splitsum{\left(\itersum{\const C1}\right)}{\splitsum{\const S0}{\itersum{\left(\choice{\const C1}{\const S0}\right)}}}$
-- is defined provided at least one survey is filled out, and in that
case, charges $\$1$ for each cup. The expression $\coffee=\splitsum{\itersum{\left(\splitsum m{\const{\#}0}\right)}}m$
maps the entire purchase history of the customer to the amount he
needs to pay the store.
\end{example}
\begin{example}
\label{ex:Combinators:Iteration:Swap} Let $\Sigma=\roset{a,b,\#}$,
and consider the function $\swapfn$ which maps strings of the form
$\strcat{\sigma}{\strcat{\#}{\tau}}$ where $\sigma,\tau\in\kstar{\roset{a,b}}$
to $\strcat{\tau}{\strcat{\#}{\sigma}}$. Such a function could be
used to transform names from the first-name-last-name format to the
last-name-first-name format. $\swapfn$ can be expressed by the function
expression $\repsum{\left(\splitsum{\const{\kstar{\roset{a,b}}\#}{\strempty}}{\itersum{\left(\choice{\trivconst a}{\trivconst b}\right)}}\right)}{\repsum{\const{\kstar{\Sigma}}{\#}}{\left(\splitsum{\itersum{\left(\choice{\trivconst a}{\trivconst b}\right)}}{\const{\#\kstar{\roset{a,b}}}{\strempty}}\right)}}$.
The first subexpression skips the first part of the string -- $\const{\kstar{\roset{a,b}}\#}{\strempty}$
-- and echoes the second part -- $\itersum{\left(\choice{\trivconst a}{\trivconst b}\right)}$.
The second subexpression $\const{\kstar{\Sigma}}{\#}$ inserts the
$\#$ in the middle. The third subexpression is similar to the first,
echoing the first part of the string and skipping the rest.\end{example}

\begin{example}
\label{ex:Combinators:Iteration:Strip} With $\Sigma=\roset{a,b,\#}$,
consider the function $\stripfn$ which map strings of the form $\strcat{\sigma_{1}}{\strcat{\#}{\strcat{\sigma_{2}}{\strcat[\ldots]{\#}{\sigma_{n}}}}}$
where $\sigma_{i}\in\kstar{\roset{a,b}}$ for each $i$ to $\strcat{\sigma_{1}}{\strcat{\#}{\strcat{\sigma_{2}}{\strcat[\ldots]{\#}{\sigma_{n-1}}}}}$.
This function could be used, for example, to locate the directory
of a file given its full path, or in processing website URLs. This
function is represented by the expression $\splitsum{\idfn}{\const{\#\kstar{\roset{a,b}}}{\strempty}}$.
\end{example}
From the appropriate definitions, we have:
\begin{prop}
\label{prop:Combinators:DoubleRev} Over all monoids $\tuple{\D,+,0}$,
the following identity holds: $\funcapptrad{\litersum f}{\sigma}=\funcapptrad{\funrev{\left(\itersum{\left(\funrev f\right)}\right)}}{\sigma}$.\end{prop}

\subsection{Chained sum \label{sub:Combinators:Chained-sum}}

Let $L\subseteq\kstar{\Sigma}$ be a language, and $f$ be a cost
function over $\kstar{\Sigma}$. If there exists a unique decomposition
$\sigma=\sigma_{1}\sigma_{2}\ldots\sigma_{k}$ such that $k\geq2$
and for each $i$, $\sigma_{i}\in L$, then the \emph{chained sum}
$\funcapptrad{\chainsum fL}{\sigma}=\funcapptrad f{\sigma_{1}\sigma_{2}}+\funcapptrad f{\sigma_{2}\sigma_{3}}+\cdots+\funcapptrad f{\sigma_{k-1}\sigma_{k}}$.
Otherwise, $\funcapptrad{\chainsum fL}{\sigma}=\bot$. Similarly,
if there exist unique strings $\sigma_{1}$, $\sigma_{2}$, \ldots{},
$\sigma_{k}$ such that $k\geq2$ and for all $i$, $\sigma_{i}\in L$,
then the \emph{left-chained sum} $\funcapptrad{\lchainsum fL}{\sigma}=\funcapptrad f{\sigma_{k-1}\sigma_{k}}+\funcapptrad f{\sigma_{k-2}\sigma_{k-1}}+\cdots+\funcapptrad f{\sigma_{1}\sigma_{2}}$.
Otherwise, $\funcapptrad{\lchainsum fL}{\sigma}=\bot$.
\begin{example}
\label{ex:Combinators:Chained-sum:Shuffle} Let $\Sigma=\roset{a,b}$
and let $\func{\shuffleexfn}{\kstar{\Sigma}}{\kstar{\Sigma}}$ be
the following function: for $\sigma=a^{m_{1}}ba^{m_{2}}b\ldots a^{m_{k}}b$,
with $k\geq2$, $\funcapptrad{\shuffleexfn}{\sigma}=a^{m_{2}}b^{m_{1}}a^{m_{3}}b^{m_{2}}\ldots a^{m_{k}}b^{m_{k-1}}$,
and for all other $\sigma$, $\funcapptrad{\shuffleexfn}{\sigma}=\bot$.
See figure \ref{fig:Combinators:Chained-sum:Shuffle:Defn}.

\begin{figure*}
\begin{centering}
\subfloat[\label{fig:Combinators:Chained-sum:Shuffle:Defn} Definition of $\ensuremath{\funcapptrad{\shuffleexfn}{\sigma}}$.]{\begin{tikzpicture}[node distance=0.1]

  \node                    (out)    {$\funcapptrad{\shuffleexfn}{\sigma}$:};
  \node [right=of out]     (oa2)    {$a^{m_2}$};
  \node [right=of oa2]     (ob1)    {$b^{m_1}$};
  \node [right=of ob1]     (oa3)    {$a^{m_3}$};
  \node [right=of oa3]     (ob2)    {$b^{m_2}$};

  \node [above=0.7 of out]   (sigma) {$\sigma$:};
  \node [above=0.7 of oa2]   (a1)    {$a^{m_1}$};
  \node [above=0.7 of ob1]   (b1)    {$b$};
  \node [above=0.7 of oa3]   (a2)    {$a^{m_2}$};
  \node [above=0.7 of ob2]   (b2)    {$b$};
  \node [right=of b2]      (a3)    {$a^{m_3}$};
  \node [right=of a3]      (b3)    {$b$};
  \node [right=of b3]      (idots) {$\ldots$};
  \node [right=of idots]   (al)    {$a^{m_{k - 1}}$};
  \node [right=of al]      (bl)    {$b$};
  \node [right=of bl]      (ak)    {$a^{m_k}$};
  \node [right=of ak]      (bk)    {$b$};

  \node [below=0.7 of idots] (odots)  {$\ldots$};
  \node [below=0.7 of al]    (oak)    {$a^{m_k}$};
  \node [below=0.7 of bl]    (obl)    {$b^{m_{k - 1}}$};

  \path [->] (a1) edge node {} (ob1);
  \path [->] (a2) edge node {} (oa2);
  \path [->] (a3) edge node {} (oa3);
  \path [->] (a2) edge node {} (ob2);
  \path [->] (ak) edge node {} (oak);
  \path [->] (al) edge node {} (obl);

\end{tikzpicture}

}
\par\end{centering}

\begin{centering}
\subfloat[\label{fig:Combinators:Chained-sum:Shuffle:Expr} Each patch $P_{i}$
is a string of the form $\kstar ab$.]{\begin{tikzpicture}[node distance=0.1]

  \node                  (sigma) {$\sigma$:};

  \node [right=of sigma] (P1)   {$P_{1}$};
  \node [below=0.7 of P1]  (P1a)  {$P_{1}$};
  \node [right=of P1a]   (P1b)  {$P_{1}$};

  \node [right=of P1b]   (P2a)  {$P_{2}$};
  \node [above=0.7 of P2a] (P2)   {$P_{2}$};
  \node [right=of P2a]   (P2b)  {$P_{2}$};

  \node [right=of P2b]   (P3a)  {$P_{3}$};
  \node [above=0.7 of P3a] (P3)   {$P_{3}$};
  \node [right=of P3a]   (P3b)  {$P_{3}$};

  \node [right=of P3b]   (Pda)  {$\ldots$};
  \node [above=0.7 of Pda] (Pd)   {$\ldots$};

  \node [right=of Pda]   (Pla)  {$P_{k - 1}$};
  \node [above=0.7 of Pla] (Pl)   {$P_{k - 1}$};
  \node [right=of Pla]   (Plb)  {$P_{k - 1}$};

  \node [right=of Plb]   (Pka)  {$P_{k}$};
  \node [above=0.7 of Pka] (Pk)   {$P_{k}$};
  \node [right=of Pka]   (Pkb)  {$P_{k}$};

  \path [->] (P1) edge node {} (P1a);
  \path [->] (P1) edge node {} (P1b);
  \path [->] (P2) edge node {} (P2a);
  \path [->] (P2) edge node {} (P2b);
  \path [->] (P3) edge node {} (P3a);
  \path [->] (P3) edge node {} (P3b);
  \path [->] (Pl) edge node {} (Pla);
  \path [->] (Pl) edge node {} (Plb);
  \path [->] (Pk) edge node {} (Pka);
  \path [->] (Pk) edge node {} (Pkb);

  \draw [decorate, decoration={brace, mirror}] (P1b.south west) -- coordinate[midway](P12) (P2a.south east);
  \draw [decorate, decoration={brace, mirror}] (P2b.south west) -- coordinate[midway](P23) (P3a.south east);
  \draw [decorate, decoration={brace, mirror}] (Plb.south west) -- coordinate[midway](Plk) (Pka.south east);

  \node [below=0.005 of P12] (P12g) {};
  \node [below=0.005 of P23] (P23g) {};
  \node [below=0.005 of Plk] (Plkg) {};

  \node [below=0.7 of P12] (fP12) {$\funcapptrad{f}{P_{1}, P_{2}}$};
  \node [below=0.7 of P23] (fP23) {$\funcapptrad{f}{P_{2}, P_{3}}$};
  \node [below=0.7 of Plk] (fPlk) {$\funcapptrad{f}{P_{k - 1}, P_{k}}$};

  \path [->] (P12g) edge node {} (fP12);
  \path [->] (P23g) edge node {} (fP23);
  \path [->] (Plkg) edge node {} (fPlk);

\end{tikzpicture}

}
\par\end{centering}

\caption{\label{fig:Combinators:Chained-sum:Shuffle} Defining and expressing
$\ensuremath{\funcapptrad{\shuffleexfn}{\sigma}}$ using function
combinators.}
\end{figure*}

We first divide $\sigma$ into chunks of text $P_{i}$, each of the
form $\kstar ab$. Similarly the output may also be divided into patches,
$P_{i}^{\prime}$. Each input patch $P_{i}$ should be scanned twice,
first to produce the $a$-s to produce $P_{i-1}^{\prime}$, and then
again to produce the $b$-s in $P_{i}^{\prime}$. Let $L=\kstar ab$
be the language of these patches. It follows that $\shuffleexfn=\chainsum fL$,
where $f=\lsplitsum{\left(\splitsum{\itersum{\const ab}}{\const b{\strempty}}\right)}{\left(\splitsum{\itersum{\const aa}}{\const b{\strempty}}\right)}$.
\end{example}
The motivation behind the chained sum is two-fold: first, we believe
that $\shuffleexfn$ is inexpressible using the remaining operators,
and second, the operation naturally emerges as an idiom during the
proof of theorem \ref{thm:Noncomm}.

\subsection{Function composition \label{sub:Combinators:Composition}}

Let $\func f{\kstar{\Sigma}}{\kstar{\Gamma}_{\bot}}$ and $\func g{\kstar{\Gamma}}{\D}$
be two cost functions. The \emph{composition $\comp gf$} is defined
as $\funcapptrad{\comp gf}{\sigma}=\funcapptrad g{\funcapptrad f{\sigma}}$,
if $\funcapptrad f{\sigma}$ and $\funcapptrad g{\funcapptrad f{\sigma}}$
are defined, and $\funcapptrad{\comp gf}{\sigma}=\bot$ otherwise.
\begin{example}
\label{ex:Combinators:Composition:Shuffle} Composition is an alternative
to chained sum for expressive completeness. Let $\autobox{\mathit{copy}}_{L}=\repsum{\left(\splitsum{\itersum{\trivconst a}}b\right)}{\left(\splitsum{\itersum{\trivconst a}}b\right)}$
be the function which accepts strings from $L$ and repeats them twice.
The first step of the transformation is therefore the expression $\itersum{\autobox{\mathit{copy}}_{L}}$.
We then drop the first copy of $P_{1}$ and the last copy of $P_{k}$
-- this is achieved by the expression $\autobox{\mathit{drop}}=\splitsum{\const L{\strempty}}{\splitsum{\idfn}{\const L{\strempty}}}$.
The function $\autobox{\mathit{ensurelen}}=\repsum{\idfn}{\const{\kplus{\Sigma}}{\strempty}}$
echoes its input, but also ensures that the input string contains
at least two patches. The final step is to specify the function $f$
which examines pairs of adjacent patches, and first echoes the $a$-s
from the second patch, and then transforms the $a$-s from the first
patch into $b$-s. $f=\lsplitsum{\left(\splitsum{\itersum{\const ab}}{\const b{\strempty}}\right)}{\left(\splitsum{\itersum{\const aa}}{\const b{\strempty}}\right)}$.
Thus, $\shuffleexfn=\comp f{\comp{\autobox{\mathit{ensurelen}}}{\comp{\autobox{\mathit{drop}}}{\itersum{\autobox{\mathit{copy}}_{L}}}}}$.
\end{example}
Observe that the approach in example \ref{ex:Combinators:Composition:Shuffle}
can be used to express the chained sum operation itself in terms of
composition. Pick a symbol $@\notin\Sigma$, and extend $f$ to $\arrow{\kstar{\left(\union{\Sigma}{\roset @}\right)}}{\D}$
by defining $\funcapptrad f{\sigma}=\bot$ whenever $\sigma$ contains
an occurrence of $@$. Let $\idfn$ be the identity function for strings
over $\Sigma$, and $\autobox{\mathit{copy}}_{L}$ be that function
which maps strings $\sigma\in L$ to $\sigma@\sigma@$, and undefined
otherwise. $\autobox{\mathit{copy}}_{L}=\repsum{\left(\splitsum{\idfn}{\const{\strempty}@}\right)}{\left(\splitsum{\idfn}{\const{\strempty}@}\right)}$.
Let $\autobox{\mathit{drop}}_{L}$ be $\splitsum{\const{L@}{\strempty}}{\splitsum{\itersum{\left(\splitsum{\idfn}{\splitsum{\const @{\strempty}}{\splitsum{\idfn}{\const @@}}}\right)}}{\const{L@}{\strempty}}}$.
Therefore, given a string $\sigma$ uniquely decomposed as $\sigma=\sigma_{1}\sigma_{2}\ldots\sigma_{k}$,
where for each $i$, $\sigma_{i}\in L$, $\comp{\autobox{\mathit{drop}}_{L}}{\itersum{\autobox{\mathit{copy}}_{L}}}$
maps it to $\strcat{\sigma_{1}}{\strcat{\sigma_{2}}{\strcat @{\strcat{\sigma_{2}}{\strcat{\sigma_{3}}{\strcat[\ldots]@{\strcat{\sigma_{k-1}}{\strcat{\sigma_{k}}@}}}}}}}$.
We then have the following:
\begin{prop}
\label{prop:Combinators:Composition:Comp} For each cost function
$f$, language $L\subseteq\kstar{\Sigma}$, and string $\sigma\in\kstar{\Sigma}$,
\begin{enumerate}
\item $\funcapptrad{\chainsum fL}{\sigma}=\funcapptrad{\comp{\itersum{\left(\splitsum f{\const @{\strempty}}\right)}}{\comp{\autobox{\mathit{ensurelen}}}{\comp{\autobox{\mathit{drop}}_{L}}{\itersum{\autobox{\mathit{copy}}_{L}}}}}}{\sigma}$,
and
\item $\funcapptrad{\lchainsum fL}{\sigma}=\funcapptrad{\comp{\litersum{\left(\splitsum f{\const @{\strempty}}\right)}}{\comp{\autobox{\mathit{ensurelen}}}{\comp{\autobox{\mathit{drop}}_{L}}{\itersum{\autobox{\mathit{copy}}_{L}}}}}}{\sigma}$.\end{enumerate}
\end{prop}

\section{Regular Functions are Closed under Combinators \label{sec:Combinators-to-RegFuns}}

As mentioned in the introduction, there are multiple equivalent definitions
of regular functions. In this paper, we will use the operational model
of copyless cost register automata ($\ccrasimp$) as the yardstick
for regularity. A $\ccrasimp$ is a finite state machine which makes
a single left-to-right pass over the input string. It maintains a
set of registers which are updated on each transition. Examples of
register updates include $v:=u+v+d$ and $v:=d+v$, where $d\in\D$
is a constant. The important restrictions are that transitions and
updates are test-free -- we do not permit conditions such as ``$q$
goes to $q^{\prime}$ on input $a$, provided $v\geq5$'' -- and
that the update expressions satisfy the copyless (or single-use) requirement.
$\ccrasimp$s are a generalization of streaming string transducers
to arbitrary monoids. The goal of this paper is to show that functions
expressible using the combinators introduced in section \ref{sec:Combinators}
are exactly the class of regular functions. In this section, we formally
define $\ccrasimp$s, and show that every function expression represents
a regular function.

\subsection{Cost register automata \label{sub:Combinators-to-RegFuns:CCRA}}
\begin{defn}
\label{defn:Combinators-to-RegFuns:Copyless} Let $V$ be a finite
set of registers. We call a function $\func fV{\kstar{\left(\union V{\D}\right)}}$
\emph{copyless} if the following two conditions hold:
\begin{enumerate}
\item For all registers $u,v\in V$, $v$ occurs at most once in $\funcapptrad fu$,
and
\item for all registers $u,v,w\in V$, if $u\neq w$ and $v$ occurs in
$\funcapptrad fu$, then $v$ does not occur in $\funcapptrad fw$.
\end{enumerate}
Similarly, a string $e\in\kstar{\left(\union V{\D}\right)}$ is copyless
if each register $v$ occurs at most once in $e$.
\end{defn}

\begin{defn}[Copyless $\crasimp$ \cite{CRA-LICS}]
 \label{defn:Combinators-to-RegFuns:CCRA} A \emph{$\ccrasimp$}
is a tuple $M=\tuple{Q,\Sigma,V,\delta,\mu,q_{0},F,\nu}$, where 

\begin{enumerate}
\item $Q$ is a finite set of states,
\item $\Sigma$ is a finite input alphabet,
\item $V$ is a finite set of registers,
\item $\func{\delta}{\cart Q{\Sigma}}Q$ is the state transition function,
\item $\func{\mu}{\cart Q{\cart{\Sigma}V}}{\kstar{\left(\union V{\D}\right)}}$
is the register update function such that for all $q$ and $a$, the
partial application $\func{\funcapptrad{\mu}{q,a}}V{\kstar V}$ is
a copyless function over $V$,
\item $q_{0}\in Q$ is the initial state,
\item $F\subseteq Q$ is the set of final states, and
\item $\func{\nu}F{\kstar{\left(\union V{\D}\right)}}$ is the output function,
such that for all $q$, the output expression $\funcapptrad{\nu}q$
is copyless.\end{enumerate}

\end{defn}
The semantics of a $\ccrasimp$ $M$ is specified using configurations.
A \emph{configuration} is a tuple $\gamma=\tuple{q,\autobox{\mathit{val}}}$
where $q\in Q$ is the current state and $\func{\autobox{\mathit{val}}}V{\D}$
is the register valuation. The initial configuration is $\gamma_{0}=\tuple{q_{0},\autobox{\mathit{val}}_{0}}$,
where $\funcapptrad{\autobox{\mathit{val}}_{0}}v=0$, for all $v$.
For simplicity of notation, we first extend $\autobox{\mathit{val}}$
to $\arrow{\union V{\D}}{\D}$ by defining $\funcapptrad{\autobox{\mathit{val}}}d=d$,
for all $d\in\D$, and then further extend it to strings $\func{\autobox{\mathit{val}}}{\kstar{\left(\union V{\D}\right)}}{\D}$,
by defining $\funcapptrad{\autobox{\mathit{val}}}{v_{1}v_{2}\ldots v_{k}}=\funcapptrad{\autobox{\mathit{val}}}{v_{1}}+\funcapptrad{\autobox{\mathit{val}}}{v_{2}}+\cdots+\funcapptrad{\autobox{\mathit{val}}}{v_{k}}$.
If the machine is in the configuration $\gamma=\tuple{q,\autobox{\mathit{val}}}$,
then on reading the symbol $a$, it transitions to the configuration
$\gamma^{\prime}=\tuple{q^{\prime},\autobox{\mathit{val}}^{\prime}}$,
and we write $\gamma\to^{a}\gamma^{\prime}$, where $q^{\prime}=\funcapptrad{\delta}{q,a}$,
and for all $v$, $\funcapptrad{\autobox{\mathit{val}}^{\prime}}v=\funcapptrad{\autobox{\mathit{val}}}{\funcapptrad{\mu}{q,a,v}}$.

We now define the function \emph{$\func{\interp M}{\kstar{\Sigma}}{\D_{\bot}}$
computed by $M$}. On input $\sigma\in\kstar{\Sigma}$, say $\gamma_{0}\to^{\sigma}\tuple{q_{f},\autobox{\mathit{val}}_{f}}$.
If $q_{f}\in F$, then $\funcapptrad{\interp M}{\sigma}=\funcapptrad{\autobox{\mathit{val}}}{\funcapptrad{\nu}{q_{f}}}$.
Otherwise, $\funcapptrad{\interp M}{\sigma}=\bot$.

A cost function is \emph{regular} if it can be computed by a $\ccrasimp$.
A streaming string transducer is a $\ccrasimp$ where the range $\D$
is the set of strings $\kstar{\Gamma}$ over the output alphabet under
concatenation.
\begin{example}
\label{ex:Combinators-to-RegFuns:CCRA} We present an example of an
$\sst$ in figure \ref{fig:Combinators-to-RegFuns:CCRA}. The machine
computes the function $\shuffleexfn$ from example \ref{ex:Combinators:Chained-sum:Shuffle}.
It maintains $3$ registers $x$, $y$ and $z$, all initially holding
the value $\strempty$. The register $x$ holds the current output.
On viewing each $a$ in the input string, the machine commits to appending
the symbol to its output. Depending on the suffix, this $a$ may also
be used to eventually produce a $b$ in the output. This provisional
value is stored in the register $z$. The register $y$ holds the
$b$-s produced by the previous run of $a$-s while the machine is
reading the next patch of $a$-s.

\begin{figure*}
\begin{centering}
\pgfmathwidth{"$\quoset{C}{x := x + 2}$"}
\def\stringswidth{\pgfmathresult}
\pgfmathparse{2.0 * \stringswidth}
\global\edef\stringswidth{\pgfmathresult}

\begin{tikzpicture}[node distance=\stringswidth pt]

  \node [state, initial]    (q0) {$q_0$};
  \node [state, right=of q0] (q1) {$q_1$};
  \node [state with output, right=of q1] (q2) {$q_2$ \nodepart{lower} $x$};

  \path [->] (q0)
             edge [loop above] node
             {$\quoset a{\begin{array}{rcl} z & := & zb \end{array}}$}
             (q0);

  \path [->] (q0)
             edge node [above]
             {$\quoset b{\begin{array}{rcl} x & := & xy \\ y & := & z \\ z & := & \strempty \end{array}}$}
             (q1);

  \path [->] (q1)
             edge [loop above] node
             {$\quoset a{\begin{array}{rcl} x & := & xa \\ z & := & zb \end{array}}$}
             (q1);

  \path [->] (q1)
             edge [bend left] node [above]
             {$\quoset b{\begin{array}{rcl} x & := & xy \\ y & := & z \\ z & := & \strempty \end{array}}$}
             (q2);

  \path [->] (q2)
             edge [bend left] node [below]
             {$\quoset a{\begin{array}{rcl} x & := & xa \\ z & := & zb \end{array}}$}
             (q1);

  \path [->] (q2)
             edge [loop above] node
             {$\quoset b{\begin{array}{rcl} x & := & xy \\ y & := & z \\ z & := & \strempty \end{array}}$}
             (q2);

\end{tikzpicture}
\par\end{centering}

\caption{\label{fig:Combinators-to-RegFuns:CCRA} Streaming string transducer
computing $\shuffleexfn$. $q_{2}$ is the only accepting state. The
annotation ``$x$'' in state $q_{2}$ specifies the output function.
On each transition, registers whose updates are not specified are
left unchanged.}
\end{figure*}
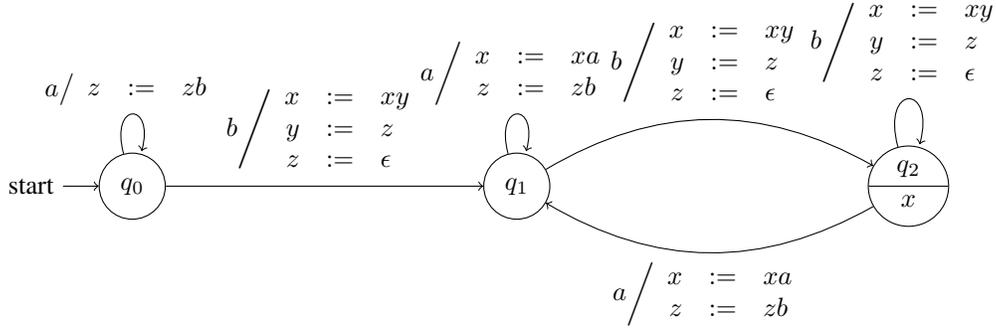

\end{example}

\subsection{Additive cost register automata \label{sub:Combinators-to-RegFuns:ACRA}}

We recall that when $\D$ is a commutative monoid, $\ccrasimp$s are
equivalent in expressiveness to the simpler model of additive cost
register automata ($\acra$). In theorem \ref{thm:Comm}, where we
show that regular functions over commutative monoids can be expressed
using the base functions over regular languages combined using the
choice, split sum and function iteration operators, we assume that
the regular function is specified as an $\acra$. These machines drop
the copyless restriction on register updates, but require that all
updates be of the form ``$u:=v+d$'', for some registers $u$ and
$v$ and some constant $d$.

\begin{defn}[Additive CRA]
 \label{defn:Combinators-to-RegFuns:ACRA} An \emph{additive cost
register automaton ($\acra$)} is a tuple $M=\tuple{Q,\Sigma,V,\delta,\mu,q_{0},F,\nu}$,
where 
\begin{enumerate}
\item $Q$ is a finite set of states,
\item $\Sigma$ is a finite input alphabet,
\item $V$ is a finite set of registers,
\item $\func{\delta}{\cart Q{\Sigma}}Q$ is the state transition function,
\item $\func{\mu}{\cart Q{\cart{\Sigma}V}}{\cart V{\D}}$ is the register
update function,
\item $q_{0}\in Q$ is the initial state,
\item $F\subseteq Q$ is the set of final states, and
\item $\func{\nu}F{\cart V{\D}}$ is the output function.
\end{enumerate}
\end{defn}
The semantics of $\acra$s are also specified using configurations.
The initial configuration $\gamma_{0}=\tuple{q_{0},\autobox{\mathit{val}}_{0}}$
maps all registers to $0$. If the machine is in a configuration $\gamma=\tuple{q,\autobox{\mathit{val}}}$,
and reads a symbol $a$, then it transitions to the configuration
$\gamma^{\prime}=\tuple{q^{\prime},\autobox{\mathit{val}}^{\prime}}$,
written as $\gamma\to^{a}\gamma^{\prime}$, where 
\begin{enumerate}
\item $q^{\prime}=\funcapptrad{\delta}{q,a}$, and
\item for each register $u$, if $\funcapptrad{\mu}{q,a,u}=\tuple{v,d}$,
then $\funcapptrad{\autobox{\mathit{val}}^{\prime}}u=\funcapptrad{\autobox{\mathit{val}}^{\prime}}v+d$.
\end{enumerate}
We then define the function $\interp M$ computed by $M$ as follows.
On input $\sigma\in\kstar{\Sigma}$, if $\gamma_{0}\to^{\sigma}\tuple{q_{f},\autobox{\mathit{val}}_{f}}$,
and $q_{f}\in F$, then $\funcapptrad{\interp M}{\sigma}=\funcapptrad{\autobox{\mathit{val}}_{f}}{\funcapptrad{\nu}{q_{f}}}$.
Otherwise, $\funcapptrad{\interp M}{\sigma}=\bot$.
\begin{example}
\label{ex:Combinators-to-RegFuns:ACRA} In figure \ref{fig:Combinators-to-RegFuns:ACRA},
we present an $\acra$ which computes the function $\coffee$ described
in example \ref{ex:Combinators:Iteration:Coffee}. In the state $q_{\lnot S}$,
the value in register $x$ tracks how much the customer owes the establishment
if he does not fill out a survey before the end of the month, and
the value in register $y$ is the amount he should pay otherwise.

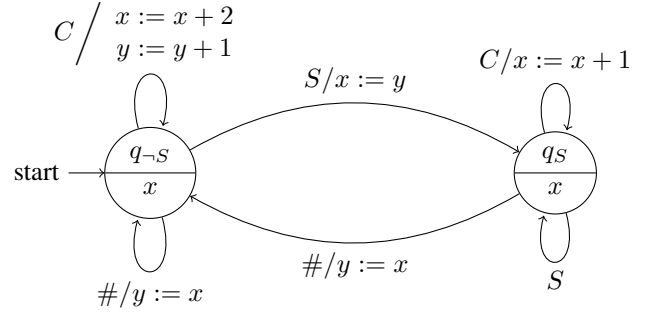
\begin{figure}
\begin{centering}
\pgfmathwidth{"$\quoset{C}{x := x + 2}$"}
\def\stringswidth{\pgfmathresult}
\pgfmathparse{2.0 * \stringswidth}
\global\edef\stringswidth{\pgfmathresult}

\begin{tikzpicture}[node distance=\stringswidth pt]

  \node [state with output, initial] (qnS) {$q_{\lnot S}$ \nodepart{lower} $x$};
  \node [state with output, right=of qnS] (qS) {$q_{S}$ \nodepart{lower} $x$};

  \path [->] (qnS) edge [loop above] node {$\quoset C {\begin{array}{c} x := x + 2\\ y := y + 1\end{array}}$} (qnS);
  \path [->] (qnS) edge [bend left, above] node {$\quoset S {x := y}$} (qS);
  \path [->] (qnS) edge [loop below] node {$\quoset \# {y := x}$} (qnS);

  \path [->] (qS) edge [loop above] node {$\quoset C {x := x + 1}$} (qS);
  \path [->] (qS) edge [loop below] node {$S$} (qS);
  \path [->] (qS) edge [bend left, below] node {$\quoset \# {y := x}$} (qnS);

\end{tikzpicture}
\par\end{centering}

\caption{\label{fig:Combinators-to-RegFuns:ACRA} $\acra$ computing $\coffee$.}
\end{figure}
\end{example}

\subsection{Regular look-ahead \label{sub:Combinators-to-RegFuns:RLA}}

An important property of regular functions is that they are closed
under \emph{regular look-ahead }\cite{SST-POPL}: a $\ccrasimp$ can
make transitions based not simply on the next symbol of the input,
but on regular properties of the as-yet-unseen suffix. To formalize
this, we introduce the notion of a \emph{look-ahead labelling}. Let
$\sigma=\strcat{\sigma_{1}}{\strcat[\ldots]{\sigma_{2}}{\sigma_{n}}}\in\kstar{\Sigma}$
be a string, and $A=\tuple{Q,\Sigma,\delta,q_{0}}$ be a DFA over
$\Sigma$. Starting in state $q_{0}$, and reading $\sigma$ in reverse,
say $A$ visits the sequence of states $q_{0}\to^{\sigma_{n}}q_{1}\to^{\sigma_{n-1}}q_{2}\to^{\sigma_{n-2}}\cdots\to^{\sigma_{1}}q_{n}$.
Then, the state of $A$ at position $i$, $q_{i}$ determines a regular
property of the suffix $\sigma_{n-i+1}\sigma_{n-i+2}\ldots\sigma_{n}$.
We term the string of states $q_{n}q_{n-1}\ldots q_{0}$ the labelling
of $\sigma$ by the \emph{look-ahead automaton $A$}.
\begin{prop}
\label{prop:Combinators-to-RegFuns:RLA-Closure} Let $A$ be a look-ahead
automaton over $\Sigma$, and let $M$ be a $\ccrasimp$ over labellings
in $\kstar Q$. Then, there is a $\ccrasimp$ machine $M^{\prime}$
over $\Sigma$, such that for every $\sigma\in\kstar{\Sigma}$, $\funcapptrad{\interp{M^{\prime}}}{\sigma}=\funcapptrad{\interp M}{\funcapptrad{\autobox{\mathit{lab}}}{\sigma}}$
where $\funcapptrad{\autobox{\mathit{lab}}}{\sigma}$ is the labelling
of $\sigma$ by $A$.\end{prop}

\subsection{From function expressions to cost register automata \label{sub:Combinators-to-RegFuns:Thm}}
\begin{thm}
\label{thm:Combinators-to-RegFuns:Thm} Every cost function expressible
using the base functions combined using the $\choiceop$, $\repsumop$,
$\splitsumop$, $\lsplitsumop$, $\itersumop$, $\litersumop$, input
reverse, composition, chained sum, and left-chained sum combinators
is regular.
\end{thm}
This can be proved by structural induction on the structure of the
function expression.  We now prove each case as a separate lemma, and these together establish
the present theorem.

\begin{lem}
\label{lem:Combinators-to-RegFuns:Base} For all regular languages
$L\subseteq\kstar{\Sigma}$, and $d\in\D$, $\const Ld$ is a regular
function.\end{lem}
\begin{IEEEproof}
Consider the DFA $A=\tuple{Q,\Sigma,\delta,q_{0},F}$ accepting $L$,
and construct the machine $M=\tuple{Q,\Sigma,\emptyset,\delta,\mu,q_{0},F,\nu}$,
where $\funcapptrad{\nu}q=d$, for all $q\in F$. This machine has
the same state space as $A$, but does not maintain any registers.
In every final state, the machine outputs the constant $d\in\D$.
The domain of the register update function $\mu$ is empty, and so
we do not specify it. Clearly, $\funcapptrad{\interp M}{\sigma}=\funcapptrad{\const Ld}{\sigma}=d$,
for each $\sigma,$and it follows that $\const Ld$ is a regular function.\end{IEEEproof}
\begin{lem}
\label{lem:Combinators-to-RegFuns:ChoiceRepSum} Whenever $f$ and
$g$ are regular functions, $\choice fg$ and $\repsum fg$ are also
regular.\end{lem}
\begin{IEEEproof}
Let $f$ and $g$ be computed by the $\ccrasimp$s $M_{f}=(Q_{f},\Sigma,V_{f},\delta_{f},\mu_{f},q_{0f},F_{f},\nu_{f})$
and $M_{g}=(Q_{g},\Sigma,V_{g},\delta_{g},\mu_{g},q_{0g},F_{g},\nu_{g})$
respectively. We use the product construction to create the machines
$M_{\choice fg}$ and $M_{\repsum fg}$ that compute $\choice fg$
and $\repsum fg$ respectively. The idea is to run both machines in
parallel, and in the case of $M_{\choice fg}$, output depending on
which machines are in accepting states. In $M_{\repsum fg}$, we output
only if both machines are accepting, and then output the sum of the
outputs of both machines.

Assume, without loss of generality, that $\intersection{V_{f}}{V_{g}}=\emptyset$.
Define $M_{\choice fg}=(\cart{Q_{f}}{Q_{g}},\Sigma,\union{V_{f}}{V_{g}},\delta,\mu,(q_{0f},q_{0g}),F_{\choice fg},\nu_{\choice fg})$
and $M_{\repsum fg}=(\cart{Q_{f}}{Q_{g}},\Sigma,\union{V_{f}}{V_{g}},\delta,\mu,(q_{0f},q_{0g}),F_{\repsum fg},\nu_{\repsum fg})$,
where 
\begin{enumerate}
\item for each $q_{1}$, $q_{2}$ and $a$, $\delta((q_{1},q_{2}),a)=(\delta_{f}(q_{1},a),\delta_{g}(q_{2},a))$,
\item if $v\in V_{f}$, then $\funcapptrad{\mu}{\tuple{q_{1},q_{2}},a,v}=\funcapptrad{\mu_{f}}{q_{1},a,v}$,
and otherwise, $\funcapptrad{\mu}{\tuple{q_{1},q_{2}},a,v}=\funcapptrad{\mu_{g}}{q_{2},a,v}$,
\item $F_{\choice fg}=\union{\cart{F_{f}}{Q_{g}}}{\cart{Q_{f}}{F_{g}}}$,
and $F_{\repsum fg}=\cart{F_{f}}{F_{g}}$,
\item for all $\tuple{q_{1},q_{2}}\in F_{\choice fg}$, if $q_{1}\in F_{f}$,
then $\funcapptrad{\nu}{q_{1},q_{2}}=\funcapptrad{\nu_{f}}{q_{1}}$,
and otherwise $\funcapptrad{\nu}{q_{1},q_{2}}=\funcapptrad{\nu_{g}}{q_{2}}$,
and
\item for all $\tuple{q_{1},q_{2}}\in F_{\repsum fg}$, $\funcapptrad{\nu}{q_{1},q_{2}}=\funcapptrad{\nu_{f}}{q_{1}}+\funcapptrad{\nu_{g}}{q_{2}}$.
\end{enumerate}
Since the sets of registers are disjoint, observe that the register
updates and output functions just defined are copyless. It follows
that $M_{\choice fg}$ and $M_{\repsum fg}$ compute $\choice fg$
and $\repsum fg$ respectively.\end{IEEEproof}
\begin{lem}
\label{lem:Combinators-to-RegFuns:SplitSum} Whenever $f$ and $g$
are regular functions, $\splitsum fg$ and $\lsplitsum fg$ are also
regular.\end{lem}
\begin{IEEEproof}
Let $f$ and $g$ be computed by the $\ccrasimp$s $M_{f}=(Q_{f},\Sigma,V_{f},\delta_{f},\mu_{f},q_{0f},F_{f},\nu_{f})$
and $M_{g}=(Q_{g},\Sigma,V_{g},\delta_{g},\mu_{g},q_{0g},F_{g},\nu_{g})$
respectively. We recall that the domain $L\subseteq\kstar{\Sigma}$
over which a regular function is defined is a regular language. Let
$L_{f}$ and $L_{g}$ be the domains of $f$ and $g$ respectively.
The idea is to use regular lookahead and execute $M_{f}$ on the prefix
$\sigma_{1}\in L_{f}$, and when the lookahead automaton indicates
that the suffix $\sigma_{2}\in L_{g}$, we switch to executing $M_{g}$,
and combine the results in the output function.

Let $A_{1}$ be a lookahead automaton with state space $\union{\Sigma}{\roset{q_{01}}}$,
so that the state of $A_{1}$ indicates the next symbol of the input.
Let $A_{2}$ be a lookahead automaton which accepts strings $\sigma$
such that $\strrev{\sigma}\in L_{g}$, and let $F_{2}$ be the set
of its accepting states. The combined lookahead automaton is the product
$\cart{A_{1}}{A_{2}}$, such that the state $\tuple{a,q}$ of this
product indicates the next symbol in the input, and depending on whether
$q\in F_{2}$, whether the suffix $\sigma_{2}\in L_{g}$.

Let $A_{3}$ (with accepting states $F_{3}$) be a DFA, which on input
$\sigma$, determines whether $\sigma$ can be unambiguously split
as $\sigma=\sigma_{1}\sigma_{2}$, with $\sigma_{1}\in L_{f}$ and
$\sigma_{2}\in L_{g}$. Construct the machine $M=\brtuple{\cart{\p{\union{Q_{f}}{Q_{g}}}}{A_{3}},\cart{Q_{1}}{Q_{2}},\union{V_{f}}{\union{V_{g}}{\roset{\autobox{\mathit{total}}}}},\delta,\mu,\brtuple{q_{0f},q_{03}},\cart{F_{g}}{F_{3}},\nu}$,
where $\delta$, $\mu$, and $\nu$ operate as follows:
\begin{enumerate}
\item In a state $\tuple{q_{1},q_{3}}\in\cart{Q_{f}}{A_{3}}$, on reading
the input symbol $\tuple{a,q_{l}}$, where $q_{1}\notin F_{f}$ or
$q_{l}\notin F_{2}$, the machine transitions to $\brtuple{\funcapptrad{\delta_{f}}{q_{1},a},\funcapptrad{\delta_{3}}{q_{3},a}}$.
The registers of $M_{f}$ are updated according to $\mu_{f}$, and
the other registers are left unchanged.
\item In a state $\tuple{q_{1},q_{3}}\in\cart{Q_{f}}{A_{3}}$, on reading
the input symbol $\tuple{a,q_{l}}$, where $q_{1}\in F_{f}$ and $q_{l}\in F_{2}$,
the machine transitions to $\brtuple{q_{0g},\funcapptrad{\delta_{3}}{q_{3},a}}$.
The machine stores the output of $M_{f}$ in the register $\autobox{\mathit{total}}$,
and the other registers are left unchanged.
\item In the state $\tuple{q_{2},q_{3}}\in\cart{Q_{g}}{A_{3}}$, on reading
the input symbol $\tuple{a,q_{l}}$, the machine transitions to $\brtuple{\funcapptrad{\delta_{g}}{q_{2},a},\funcapptrad{\delta_{3}}{q_{3},a}}$.
The registers of $M_{g}$ are updated according to $\mu_{g}$, and
the other registers are left unchanged.
\item In the final state $\tuple{q_{2},q_{3f}}\in\cart{F_{g}}{F_{3}}$,
the machine outputs the value $\autobox{\mathit{total}}+\funcapptrad{\nu}{q_{2}}$.
\end{enumerate}
The machine $M$ just constructed computes the function $\splitsum fg$
using regular lookahead, and it follows that $\splitsum fg$ is regular.
Similarly, it can be shown that $\lsplitsum fg$ is also regular.
\end{IEEEproof}
Along similar lines, we have:
\begin{lem}
\label{lem:Combinators-to-RegFuns:IterSum} Whenever $f$ is a regular
function, $\itersum f$ and $\litersum f$ are also regular.\end{lem}
\begin{IEEEproof}
The main difference between this and the construction of lemma \ref{lem:Combinators-to-RegFuns:SplitSum}
are the following: the state space $Q$ of $M$ is defined as $Q=\cart{Q_{f}}{A_{3}}$,
since there is only one $\ccrasimp$ $M_{f}$. The set of registers
is $V=\union V{\roset{\autobox{\mathit{total}}}}$, and the accepting
states $F=\cart{F_{f}}{F_{3}}$.

In a state $\tuple{q_{1},q_{3}}\in\cart{Q_{f}}A$, on reading the
input symbol $\tuple{a,q_{l}}$, where $q_{1}\in F_{f}$, and $q_{l}\in F_{2}$,
the machine transitions back to $\tuple{q_{0f},\funcapptrad{\delta_{3}}{q_{3},a}}$.
The machine appends the output of $M_{f}$ to the right of the register
$\autobox{\mathit{total}}$, and all other registers are cleared to
$0$.

The machine thus constructed computes $\itersum f$. If the machine
were to append the output of $M_{f}$ to the left of $\autobox{\mathit{total}}$,
then it would compute $\litersum f$. Thus, both function expressions
are regular.
\end{IEEEproof}
The next lemma was first proved in \cite{CRA-LICS}. It can also be
seen as a consequence of lemma \ref{lem:Combinators-to-RegFuns:Comp},
because for all $\sigma$, $\funcapptrad{\funrev f}{\sigma}=\funcapptrad{\comp f{\autobox{\mathit{reverse}}}}{\sigma}$,
where $\autobox{\mathit{reverse}}=\litersum{\choiceop\ruset{\const aa}{a\in\Sigma}}$
is the function which reverses its input.
\begin{lem}
\label{lem:Combinators-to-RegFuns:FunRev} Whenever $f$ is a regular
function, so is $\funrev f$.
\end{lem}

\begin{lem}
\label{lem:Combinators-to-RegFuns:Comp} Whenever $\func f{\kstar{\Gamma}}{\D}$
and $\func g{\kstar{\Sigma}}{\kstar{\Gamma}}$ are regular functions,
$\comp fg$ is also a regular function.\end{lem}
\begin{IEEEproof}
Since $\sst$s are closed under composition, if $\func f{\kstar{\Gamma}}{\D}$
and $\func g{\kstar{\Sigma}}{\kstar{\Gamma}}$ are regular functions,
it follows that $\comp fg$ is also a regular function.\end{IEEEproof}
\begin{lem}
\label{lem:Combinators-to-RegFuns:ChainedSum} Whenever $f$ is a
regular function, and $L\subseteq\kstar{\Sigma}$ is a regular language,
$\chainsum fL$ and $\lchainsum fL$ are also regular functions.\end{lem}
\begin{IEEEproof}
From proposition \ref{prop:Combinators:Composition:Comp} and lemma
\ref{lem:Combinators-to-RegFuns:Comp}.
\end{IEEEproof}
This completes the proof of theorem \ref{thm:Combinators-to-RegFuns:Thm}.

\section{Completeness of Combinators for Commutative Monoids \label{sec:Comm}}

\begin{figure*}
\begin{centering}
\subfloat[\label{fig:Comm:ACRA}]{\begin{centering}
\begin{tikzpicture}

  \node [state with output, initial] (q0) {$q_{0}$ \nodepart{lower} $x$};

  \path [->] (q0)
             edge [loop right] node [right]
             {$\quoset a{\begin{array}{rcl}x & := & x + 1\\y & := & y + 1\end{array}}$}
             (q0);

  \path [->] (q0)
             edge [loop above] node [above]
             {$\quoset b{\begin{array}{rcl}x & := & x\\y & := & y + 1\end{array}}$}
             (q0);

  \path [->] (q0)
             edge [loop below] node [below]
             {$\quoset e{\begin{array}{rcl}x & := & y + 1\\y & := & y + 1\end{array}}$}
             (q0);

\end{tikzpicture}
\par\end{centering}

} \hfill{} \subfloat[\label{fig:Comm:NFA}]{\begin{centering}
\begin{tikzpicture}

  \node [state, accepting] (q0x) {$\tuple{q_{0}, x}$};
  \node [state, right=of q0x] (q0y) {$\tuple{q_{0}, y}$};

  \path [->] (q0x) edge [loop above] node [above] {$a_{1}$} (q0x);
  \path [->] (q0x) edge [loop left] node [left] {$b_{0}$} (q0x);

  \path [->] (q0y) edge [loop above] node [above] {$a_{1}$} (q0y);
  \path [->] (q0y) edge [loop right] node [right] {$b_{1}$} (q0y);
  \path [->] (q0y) edge [loop below] node [below] {$e_{1}$} (q0y);
  \path [->] (q0y) edge node [above] {$e_{1}$} (q0x);

\end{tikzpicture}
\par\end{centering}

}
\par\end{centering}

\caption{\label{fig:Comm} Translating an $\acra$ to the commutative calculus.
The machine operates over the alphabet $\Sigma=\roset{a,b,e}$, and
when given a string $\sigma=\sigma_{1}e\sigma_{2}e\ldots\sigma_{k}$,
where each $\sigma_{i}\in\kstar{\roset{a,b}}$, it counts the number
of $a$-s and $e$-s, but only counts those $b$-s which occur before
the final $e$. Figure \ref{fig:Comm:NFA} is the NFA that results
from the construction of theorem \ref{thm:Comm}. Both states in the
NFA are initial.}
\end{figure*}
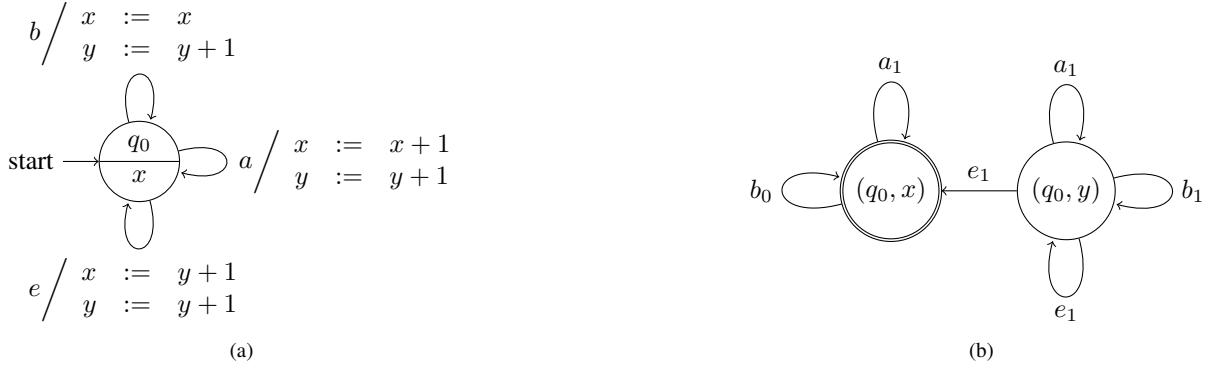

In this section, we show that if $\D$ is a commutative monoid, then
constant functions combined using the choice, split sum and function
iteration are expressively equivalent to the class of regular functions.
Consider the $\acra$ $M$ shown in figure \ref{fig:Comm:ACRA}. The
idea is to view $M$ as a non-deterministic automaton $A$ over the
set of vertices $\cart QV$: for every path $\pi=q_{0}\to^{\sigma_{1}}q_{1}\to^{\sigma_{2}}\cdots\to^{\sigma_{n}}q_{n}$
through the $\acra$, there is a corresponding path through $A$,
$\pi_{A}=\tuple{q_{0},v_{0}}\to^{\sigma_{1}}\tuple{q_{1},v_{1}}\to^{\sigma_{2}}\cdots\to^{\sigma_{n}}\tuple{q_{n},v_{n}}$,
where $v_{n}$ is the register which is output in the final state
$q_{n}$, and at each position $i$, $v_{i}$ indicates the register
whose current value flows into the final value of $v_{n}$. Observe
that this NFA $A$ is unambiguous -- for every string $\sigma$ that
is accepted by $A$, there is a unique accepting path. Furthermore,
the final value of register $v_{n}$ is simply the sum of the increments
accumulated along each transition of this accepting path. Therefore,
if the label $\tuple{q,v}\to^{a_{d}}\tuple{q^{\prime},v^{\prime}}$
along each edge is also annotated with the increment value $d$, so
that the update expression reads $\funcapptrad{\mu}{q,a,v^{\prime}}=v+d$,
then the regular expression for the language accepted $A$ -- $\kstar{\left(a_{1}+b_{0}\right)}+\kstar{\left(a_{1}+b_{1}+e_{1}\right)}e_{1}\kstar{\left(a_{1}+b_{0}\right)}$
-- can be alternatively viewed as a function expression for $\interp M$
-- $\choice{\itersum{\p{\const b0\choiceop\const a1}}}{\p{\itersum{\p{\const b1\choiceop\const a1\choiceop\const e1}}\splitsumop e_{1}\splitsumop\itersum{\p{\const b0\choiceop\const a1}}}}$.

\begin{thm}
\label{thm:Comm} If $\tuple{\D,+,0}$ is a commutative monoid, then
every regular function $\func f{\kstar{\Sigma}}{\D}$ can be expressed
using the base functions combined with the choice, split sum and iterated
sum operators.
\end{thm}
\begin{IEEEproof}
We need to show an arbitrary $\acra$ $M=\tuple{Q,\Sigma,V,\delta,\mu,q_{0},F,\nu}$
can be expressed by these combinators.

We construct an NFA $A$ with states $\cart QV$ and an alphabet $\Gamma$
consisting of a finite subset of $\cart{\Sigma}{\D}$ which are those
elements $\tuple{a,d}$ such that for some state $q\in Q$ and two
registers $v,v^{\prime}\in V$ there is an update $\funcapptrad{\mu}{q,a,v}=v^{\prime}+d$.
We will denote $\tuple{a,d}$ as $a_{d}$.

We define the transition relation as follows: $\tuple{q^{\prime},v^{\prime}}\in\funcapptrad{\delta^{\prime}}{\tuple{q,v},a_{d}}$
iff $\funcapptrad{\mu}{q,a,v^{\prime}}=v+d$ and $\funcapptrad{\delta}{q,a}=q^{\prime}$.

Assume without loss of generality that our output function takes values
in $V$. The start states of the NFA $A$ are all states in $\cart{\roset{q_{0}}}V$
and the final states are $\ruset{\tuple{q,v}}{\funcapptrad{\nu}q=v}$.

Consider any unambiguous regular expression of strings accepted by
the NFA $A$: interpret regular expression union $\cup$ as $\choiceop$,
regular expression concatenation $\cdot$ as $\splitsumop$, Kleene-{*}
as the iterated sum $\itersum f$ and input symbols $a_{d}$ as the
constant functions $\const ad$.

It can be shown by an inductive argument that the regular expression
corresponding to paths in our NFA from $\tuple{q,v}$ to $\tuple{q^{\prime},v^{\prime}}$,
when interpreted as a regular function $f$ is defined exactly on
those $\sigma\in\kstar{\Sigma}$ such that $\sigma$ is a path from
$q$ to $q^{\prime}$ with the effect that $v$ flows into $v^{\prime}$.
Moreover, the total effect of this path $\sigma$ is $v^{\prime}:=v+\funcapptrad f{\sigma}$
for all of these $\sigma$. It follows that a function expression
for $\interp M$ can be obtained from the union of the unambiguous
regular expressions from some state in $\cart{\roset{q_{0}}}V$ to
some state in $\ruset{\tuple{q,v}}{\funcapptrad{\nu}q=v}$.\end{IEEEproof}

\section{Completeness of Combinators for General Monoids \label{sec:Noncomm}}

In this section, we describe an algorithm to express every regular
function $\func f{\kstar{\Sigma}}{\D}$ as a function expression.
To simplify the presentation, we prove theorem \ref{thm:Noncomm}
only for the case of string transductions, i.e. where $\D=\kstar{\Gamma}$,
for some finite output alphabet $\Gamma$. Note that this is sufficient
to establish the theorem in its full generality: let $\Gamma_{\D}\subseteq\D$
be the (necessarily finite) set of all constants appearing in the
textual description of $M$. $M$ can be alternatively viewed as an
$\sst$ mapping input strings in $\kstar{\Sigma}$ to output strings
in $\kstar{\Gamma_{\D}}$. The restricted version of theorem \ref{thm:Noncomm}
can then be used to convert this $\sst$ to function expression form,
which when interpreted over the original domain $\D$ represents $\interp M$.
\begin{thm}
\label{thm:Noncomm} For an arbitrary finite alphabet $\Sigma$ and
monoid $\tuple{\D,+,0}$, every regular function $\func f{\kstar{\Sigma}}{\D}$
can be expressed using the base functions combined with choice, sum,
split sum, iterated sum, chained sum, and their left-additive versions.
\end{thm}
\global\long\def\pareg#1#2#3{\funcapptrad{r^{\left(#1\right)}}{#2,#3}}

\global\long\def\paregv#1#2#3#4{\funcapptrad{\vector R_{#2}^{\left(#1\right)}}{#3,#4}}

\global\long\def\paregp#1#2#3#4{\funcapptrad{\vector R_{#2}^{\left(#1\right)}}{#3,#4}}

\global\long\def\S{\mathbb{S}}

\global\long\def\shapecat#1#2{\strcat[\cdot]{#1}{#2}}

\global\long\def\regleq{\preceq}

\global\long\def\reglt{\prec}

\global\long\def\support#1{\funcapptrad{\autobox{\mathit{supp}}}{#1}}

\global\long\def\shleq{\sqsubseteq}

\global\long\def\shlt{\sqsubset}

\global\long\def\suppeq{\sim}

\subsection{From DFAs to regular expressions: A review \label{sub:Noncomm:DFA-Regex}}

The procedure to convert a $\ccrasimp$ into a function expression
is similar to the corresponding algorithm \cite{Sipser-Intro} that
transforms a DFA $A=\brtuple{Q,\Sigma,\delta,q_{0},F}$ into an equivalent
regular expression; we will also use this algorithm in our correctness
proof -- hence this review.

Let $Q=\rosetbr{q_{1},q_{2},\ldots,q_{n}}$. For each pair of states
$q,q^{\prime}\in Q$, and for $i\in\N$, $0\leq i\leq n$, $\pareg iq{q^{\prime}}$
is the set of strings $\sigma$ from $q$ to $q^{\prime}$, while
only passing through the intermediate states $\roset{q_{1},q_{2},\ldots,q_{i}}$.
This can be inductively constructed as follows:
\begin{enumerate}
\item $\pareg 0q{q^{\prime}}=\ruset{a\in\union{\Sigma}{\roset{\strempty}}}{q\to^{a}q^{\prime}}$.
\item $\pareg{i+1}q{q^{\prime}}=\pareg iq{q^{\prime}}+\pareg iq{q_{i+1}}\kstar{\pareg i{q_{i+1}}{q_{i+1}}}\pareg i{q_{i+1}}{q^{\prime}}$.
\end{enumerate}
The language $L$ accepted by $A$ is then given by the regular expression
$\sum_{q_{f}\in F}\pareg n{q_{0}}{q_{f}}$. Note that the regular
expression thus obtained is also unambiguous.

\subsection{A theory of shapes \label{sub:Noncomm:Shapes}}

In a $\ccrasimp$ $M$, the effect of processing a string $\sigma$
starting from a state $q$ can be summarized by the pair $\brtuple{\fabr{\delta}{q,\sigma},\fabr{\mu}{q,\sigma}}$
-- $\funcapptrad{\delta}{q,\sigma}$ is the state of the machine after
processing $\sigma$, and the partial application of the register
update function $\func{\fabr{\mu}{q,\sigma}}V{\kstar{\p{\union V{\Gamma}}}}$
expresses the final values of the registers in terms of their initial
ones.

Consider the expression $\funcapptrad{\mu}{q,\sigma,u}=aubcvd$, where
$u,v\in V$ are registers, and $a,b,c,d\in\kstar{\Gamma}$ are string
constants. Because of the associative property, any update expression
can be equivalently represented -- as in $\funcapptrad{\mu}{q,\sigma,u}=aub^{\prime}vd$
where $b^{\prime}=bc$ -- so that there is at most one string constant
between consecutive registers in this update expression. The summary
for $\sigma$ therefore contains the shape $\func{S_{\sigma}}V{\kstar V}$
indicating the sequence of registers in each update expression and,
for each register $v$ and each position $k$ from $1,2,\ldots,\strlen{\funcapptrad{S_{\sigma}}v}+1$,
a string $\gamma_{k}\in\kstar{\Gamma}$ indicating the $k^{\mbox{th}}$
string constant appearing in $\funcapptrad{\mu}{q,\sigma,u}$.
\begin{defn}[Shape of a path]
 \label{defn:Noncomm:Shapes} A \emph{shape} $\func SV{\kstar V}$
is a copyless function over a finite set of registers $V$. Let $\pi=q_{1}\to^{\sigma_{1}}q_{2}\to^{\sigma_{2}}\to\cdots\to^{\sigma_{n}}q_{n+1}$
be a path through a $\ccrasimp$ $M$. The \emph{shape of the path
$\pi$} is the function $\func{S_{\pi}}V{\kstar V}$ such that for
all registers $v\in V$ , $\funcapptrad{S_{\pi}}v$ is the string
projection onto $V$ of the register update expression $\funcapptrad{\mu}{q_{1},\sigma,v}$:
$\funcapptrad{S_{\pi}}v=\funcapptrad{\pi_{V}}{\funcapptrad{\mu}{q_{1},\sigma,v}}$. 
\end{defn}
We refer to a string constant in the update expression as a \emph{patch}
in the corresponding shape. Because of the copyless restriction on
the register update function, the set of all shapes over $V$ is finite.

The following is an immediate consequence of the space of shapes being
finite:
\begin{prop}
\label{prop:Noncomm:Shapes:Regular} Let $q,q^{\prime}\in Q$ be two
states in a $\ccrasimp$ $M$, and $S$ be a shape. The set of all
strings from $q$ to $q^{\prime}$ in $M$ with shape $S$ is regular.\end{prop}

\begin{example}
It is helpful to visualize shapes as bipartite graphs (figure \ref{fig:Noncomm:Shapes}),
though this representation omits some important information about
the shape. Since the shape of a path indicates the pattern in which
register values flow during computation, an edge $u\to v$ can be
informally read as ``The value of $u$ flows into $v$''. Because
of the copyless restriction, every node on the left is connected to
at most one node on the right.

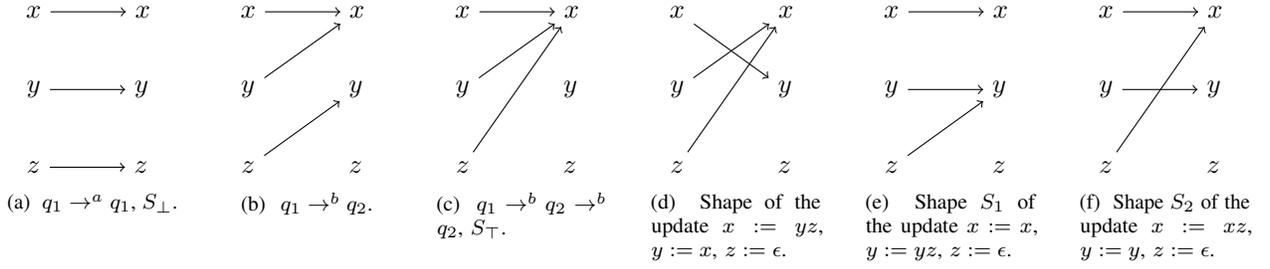
\begin{figure*}
\begin{centering}
\subfloat[\label{fig:Noncomm:Shapes:q1a} $q_{1}\to^{a}q_{1}$, $S_{\bot}$.]{\begin{centering}
\begin{tikzpicture}

  \node                 (ix) {$x$};
  \node [below=0.6 of ix] (iy) {$y$};
  \node [below=0.6 of iy] (iz) {$z$};

  \node [right=1 of ix] (ox) {$x$};
  \node [right=1 of iy] (oy) {$y$};
  \node [right=1 of iz] (oz) {$z$};

  \path [->] (ix) edge (ox);
  \path [->] (iy) edge (oy);
  \path [->] (iz) edge (oz);

\end{tikzpicture}
\par\end{centering}

} \hfill{} \subfloat[\label{fig:Noncomm:Shapes:q1b} $q_{1}\to^{b}q_{2}$.]{\begin{centering}
\begin{tikzpicture}

  \node                 (ix) {$x$};
  \node [below=0.6 of ix] (iy) {$y$};
  \node [below=0.6 of iy] (iz) {$z$};

  \node [right=1 of ix] (ox) {$x$};
  \node [right=1 of iy] (oy) {$y$};
  \node [right=1 of iz] (oz) {$z$};

  \path [->] (ix) edge (ox);
  \path [->] (iy) edge (ox);
  \path [->] (iz) edge (oy);

\end{tikzpicture}
\par\end{centering}

} \hfill{} \subfloat[\label{fig:Noncomm:Shapes:q1bb} $q_{1}\to^{b}q_{2}\to^{b}q_{2}$,
$S_{\top}$.]{\begin{centering}
\begin{tikzpicture}

  \node                 (ix) {$x$};
  \node [below=0.6 of ix] (iy) {$y$};
  \node [below=0.6 of iy] (iz) {$z$};

  \node [right=1 of ix] (ox) {$x$};
  \node [right=1 of iy] (oy) {$y$};
  \node [right=1 of iz] (oz) {$z$};

  \path [->] (ix) edge (ox);
  \path [->] (iy) edge (ox);
  \path [->] (iz) edge (ox);

\end{tikzpicture}
\par\end{centering}

} \hfill{} \subfloat[\label{fig:Noncomm:Shapes:Weird} Shape of the update $x:=yz$, $y:=x$,
$z:=\strempty$.]{\begin{centering}
\begin{tikzpicture}

  \node                 (ix) {$x$};
  \node [below=0.6 of ix] (iy) {$y$};
  \node [below=0.6 of iy] (iz) {$z$};

  \node [right=1 of ix] (ox) {$x$};
  \node [right=1 of iy] (oy) {$y$};
  \node [right=1 of iz] (oz) {$z$};

  \path [->] (ix) edge (oy);
  \path [->] (iy) edge (ox);
  \path [->] (iz) edge (ox);

\end{tikzpicture}
\par\end{centering}

} \hfill{} \subfloat[\label{fig:Noncomm:Shapes:NoncommS1} Shape $S_{1}$ of the update
$x:=x$, $y:=yz$, $z:=\strempty$.]{\begin{centering}
\begin{tikzpicture}

  \node                 (ix) {$x$};
  \node [below=0.6 of ix] (iy) {$y$};
  \node [below=0.6 of iy] (iz) {$z$};

  \node [right=1 of ix] (ox) {$x$};
  \node [right=1 of iy] (oy) {$y$};
  \node [right=1 of iz] (oz) {$z$};

  \path [->] (ix) edge (ox);
  \path [->] (iy) edge (oy);
  \path [->] (iz) edge (oy);

\end{tikzpicture}
\par\end{centering}

} \hfill{} \subfloat[\label{fig:Noncomm:Shapes:NoncommS2} Shape $S_{2}$ of the update
$x:=xz$, $y:=y$, $z:=\strempty$.]{\begin{centering}
\begin{tikzpicture}

  \node                 (ix) {$x$};
  \node [below=0.6 of ix] (iy) {$y$};
  \node [below=0.6 of iy] (iz) {$z$};

  \node [right=1 of ix] (ox) {$x$};
  \node [right=1 of iy] (oy) {$y$};
  \node [right=1 of iz] (oz) {$z$};

  \path [->] (ix) edge (ox);
  \path [->] (iy) edge (oy);
  \path [->] (iz) edge (ox);

\end{tikzpicture}
\par\end{centering}

}
\par\end{centering}

\caption{\label{fig:Noncomm:Shapes} Visualizing shapes as bipartite graphs.
Figures \ref{fig:Noncomm:Shapes:q1a}-\ref{fig:Noncomm:Shapes:q1bb}
describe the shapes of some paths in the earlier $\sst$ example of
figure \ref{fig:Combinators-to-RegFuns:CCRA}.}
\end{figure*}

\end{example}
When two paths are concatenated, their shapes are combined. We define
the \emph{concatenation $\shapecat{S_{1}}{S_{2}}$} of two shapes
$S_{1}$ and $S_{2}$ as follows. For some register $v\in V$, let
$\funcapptrad{S_{2}}v=v_{1}v_{2}\ldots v_{k}$. Then $\funcapptrad{\shapecat{S_{1}}{S_{2}}}v=s_{1}s_{2}\ldots s_{k}$,
where $s_{i}=\funcapptrad{S_{1}}{v_{i}}$. By definition, therefore, 
\begin{prop}
\label{prop:Noncomm:Shapes:Concat} Let $\pi_{1}$ and $\pi_{2}$
be two paths through a $\ccrasimp$ $M$ such that the final state
of $\pi_{1}$ is the same as the initial state of $\pi_{2}$. Then,
for all registers $v$, $\funcapptrad{S_{\pi_{1}\pi_{2}}}v=\funcapptrad{\shapecat{S_{\pi_{1}}}{S_{\pi_{2}}}}v$.
\end{prop}

\subsection{Proof outline \label{sub:Noncomm:Outline}}

To summarize the effect of a set of paths with the same shape, we
introduce the notion of an expression vector -- for a shape $S$,
an \emph{expression vector} $\vector A$ is a collection of function
expressions, such that for each register $v$, and for each patch
$k$ in $\funcapptrad Sv$, there is a corresponding function expression
$\func{\vector A_{v,k}}{\kstar{\Sigma}}{\kstar{\Gamma}}$. An expression
vector $\vector A$ \emph{summarizes a set of paths} $L$ with shape
$S$, if for each path $\pi\in L$ with initial state $q$, and input
string $\sigma$, and for each register $v$, in the update expression
$\funcapptrad{\mu}{q,\sigma,v}$, the constant value $\gamma_{k}\in\kstar{\Gamma}$
at position $k$ is given by $\funcapptrad{\vector A_{v,k}}{\sigma}$.
\begin{example}
Consider the loop $\kstar a$ at the state $q_{1}$ in the $\sst$
of figure \ref{fig:Combinators-to-RegFuns:CCRA}. Consider some concrete
string, $a^{k}$. The effect of this string is to update $x:=xa^{k}$,
$y:=y$, and $z:=zb^{k}$. The shape of this set of paths is the identity
function $\funcapptrad Sv=v$, for all $v$. Define the expression
vector $\vector A$ as follows: $\vector A_{x,1}=\vector A_{y,1}=\vector A_{y,2}=\vector A_{z,1}=\const{\kstar{\Sigma}}{\strempty}$,
$\vector A_{x,2}=\itersum{\const aa}$, and $\vector A_{z,2}=\itersum{\const ab}$.
Then $\vector A$ summarizes the set of paths $\kstar a$ at the state
$q_{1}$.
\end{example}
The outer loop of our algorithm is an iteration which proceeds in
lock-step with the DFA-to-regular expression translator. In step $i$,
for each pair of states $q,q^{\prime}\in Q$, and shape $S$, we maintain
an expression vector $\paregv iSq{q^{\prime}}$. The invariant maintained
is that $\paregv iSq{q^{\prime}}$ summarizes all paths $\sigma\in\pareg iq{q^{\prime}}$
with shape $S$.

After this iteration is complete, pick a state $q_{f}\in F$, a shape
$S$, and some register $v$. Construct the function expression $f_{S,v}=\repsum{\paregp n{S,v,1}{q_{0}}{q_{f}}}{\repsum{\paregp n{S,v,2}{q_{0}}{q_{f}}}{\repsum{\cdots}{\paregp n{S,v,\left|\funcapptrad Sv\right|+1}{q_{0}}{q_{f}}}}}$.
Because $v$ initially held the empty string $\strempty$, it follows
that for each path $q_{0}\to^{\sigma}q_{f}$ with shape $S$, the
final value in the register $v$ is given by $\funcapptrad{f_{S,v}}{\sigma}$.
We will then have constructed a function expression equivalent to
the given $\ccrasimp$ $M$.

There are therefore two steps in this construction:
\begin{enumerate}
\item Construct $\paregv 0Sq{q^{\prime}}$, for each pair of registers $q,q^{\prime}\in Q$,
and shape $S$. 
\item For each $1\leq i<n$, $0\leq j\leq i$, and for all shapes $S$,
and pairs of states $q,q^{\prime}\in Q$, given $\paregv jSq{q^{\prime}}$,
construct $\paregv{i+1}Sq{q^{\prime}}$.
\end{enumerate}

\subsection{Operations on expression vectors \label{sub:Noncomm:EVOps}}

In this subsection, we create a library of basic operations on expression
vectors, including concatenation and union.

\subsubsection{Restricting expression domains \label{sub:Noncomm:EVOps:Restrict}}

Given an expression vector $\vector A$ for a shape $S$, the domain
of the expression vector, written as $\domain{\vector A}$, is defined
as the language $\bigintersection{v,k}{\domain{\vector A_{v,k}}}$,
where $\domain{\vector A_{v,k}}$ is the domain of the component function
expressions. We would want to restrict the component expressions in
a vector so that they all have the same domain -- given a cost function
$\func f{\kstar{\Sigma}}{\kstar{\Gamma}}$ and a language $L\subseteq\kstar{\Sigma}$,
we define the \emph{restriction of $f$ to $L$} as $\restrict fL=\repsum f{\const L{\strempty}}$.
This is equivalent to saying that $\funcapptrad{\restrict fL}{\sigma}=\funcapptrad f{\sigma}$,
if $\sigma\in L$, and $\funcapptrad{\restrict fL}{\sigma}=\bot$,
otherwise. We extend this to restrict expression vectors $\vector A$
to languages $L$, $\restrict{\vector A}L$, by defining $\left(\restrict{\vector A}L\right)_{v,k}$
as $\restrict{\vector A_{v,k}}L$.

\subsubsection{Shifting expressions \label{sub:Noncomm:EVOps:Shift}}

Given a cost function $f$ and a language $L$, the \emph{left-shifted
function $\lshift fL$} is the function which reads an input string
in $\domain f\cdot L$, and applies $f$ to the prefix and ignores
the suffix, provided the split is unique, i.e. $\lshift fL=\splitsum f{\const L{\strempty}}$.
Similarly, the \emph{right-shifted function $\rshift fL=\splitsum{\const L{\strempty}}f$}.
The shift operators can also be extended to expression vectors: $\lshift{\vector A}L$
is defined as $\left(\lshift{\vector A}L\right)_{v,k}=\lshift{\restrict{\vector A_{v,k}}{\domain{\vector A}}}L$,
and $\rshift{\vector A}L$ is defined as $\left(\rshift{\vector A}L\right)_{v,k}=\rshift{\restrict{\vector A_{v,k}}{\domain{\vector A}}}L$.

\subsubsection{Concatenation \label{sub:Noncomm:EVOps:Concat}}

Let $L$ be a set of paths with shape $S$, and $L^{\prime}$ be a
set of paths with shape $S^{\prime}$. Let the expression vectors
$\vector A$ and $\vector B$ summarize paths in $L$ and $L^{\prime}$
respectively. We now construct an expression vector $\vector A\cdot\vector B$
which summarizes unambiguous paths in $L\cdot L^{\prime}$.

Consider a path $\pi\in L\cdot L^{\prime}$ which can be unambiguously
decomposed as $\pi=\pi_{1}\pi_{2}$ with $\pi_{1}\in L$ and $\pi_{2}\in L^{\prime}$.
When applying $\vector A$ (resp. $\vector B$) to this path, we should
shift the expression vector to examine only $\pi_{1}$ (resp. $\pi_{2}$).
Thus, define $\vector A^{\prime}=\lshift{\vector A}{\domain{\vector B}}$,
and $\vector B^{\prime}=\rshift{\vector B}{\domain{\vector A}}$.

Pick a register $v$, and let $v:=f_{1}v_{1}f_{2}v_{2}\ldots v_{k}f_{k+1}$
be the update expression for $v$ in $\vector B^{\prime}$. For each
register $v_{i}$ in the right-hand side, let $v_{i}:=f_{i1}v_{i1}f_{i2}v_{i2}\ldots v_{ik_{i}}f_{ik_{i}+1}$
be the update expression for $v_{i}$ in $\vector A^{\prime}$. View
string concatenation as the function combinator $+$, and substitute
the expression for each $v_{i}$ in $\vector A^{\prime}$ into the
expression for $v$ in $\vector B^{\prime}$. Then, observe that $\left(\vector A\cdot\vector B\right)_{v,k}$
is the $k^{\mbox{th}}$ function expression in the string that results.

\subsubsection{Choice \label{sub:Noncomm:EVOps:Choice}}

Let $\vector A$ and $\vector B$ be expression vectors, both for
some shape $S$. Let $\vector A^{\prime}=\restrict{\vector A}{\domain{\vector A}}$
and $\vector B^{\prime}=\restrict{\vector B}{\domain{\vector B}}$.
Then, define the choice $\choice{\vector A}{\vector B}$ is the expression
vector for shape $S$ such that for each register $v$ and patch $k$,
$\left(\choice{\vector A}{\vector B}\right)_{v,k}=\choice{\vector A_{v,k}^{\prime}}{\vector B_{v,k}^{\prime}}$.
\begin{claim}
\label{clm:Noncomm:EVOps:Choice} If $L$ and $L^{\prime}$ are disjoint
sets of paths with the same shape $S$, such that $\vector A$ summarizes
paths in $L$ and $\vector B$ summarizes paths in $L^{\prime}$,
then $\choice{\vector A}{\vector B}$ summarizes paths in $\union L{L^{\prime}}$.
\end{claim}
The notation $\choiceop\rosetbr{f_{1},f_{2},\ldots,f_{k}}$ is shorthand
for the expression $\choice{f_{1}}{\choice{f_{2}}{\choice{\cdots}{f_{k}}}}$.
We ensure that when this notation is used, the functions have mutually
disjoint domains, so the order is immaterial.

\subsection{Constructing $\paregv 0Sq{q^{\prime}}$ \label{sub:Noncomm:Rs0}}

For each string $a\in\union{\Sigma}{\roset{\strempty}}$, and each
pair of states $q,q^{\prime}\in Q$ such that $q\to^{a}q^{\prime}$,
if $S$ is the shape of the update expression of $q\to^{a}q^{\prime}$,
we define $\paregv aSq{q^{\prime}}$ as follows. For each register
$v$ and patch $k$ in $\funcapptrad Sv$, $\paregp a{S,v,k}q{q^{\prime}}=\const a{\gamma_{v,k}}$,
where $\gamma_{v,k}$ is the $k^{\mbox{th}}$ string constant appearing
in the update expression $\funcapptrad{\mu}{q,a,v}$. For all other
$a\in\union{\Sigma}{\roset{\strempty}}$, $q,q^{\prime}\in Q$, and
shapes $S$, define $\paregv aSq{q^{\prime}}=\botfn$. Finally, $\paregv 0Sq{q^{\prime}}=\mbox{ }\choiceop\rusetbr{\paregv aSq{q^{\prime}}}{a\in\union{\Sigma}{\roset{\strempty}}}$.
By construction, 
\begin{claim}
\label{clm:Noncomm:Rs0} For each pair of states $q,q^{\prime}\in Q$
and shape $S$, $\paregv 0Sq{q^{\prime}}$ summarizes all paths $\sigma\in\pareg 0q{q^{\prime}}$
from $q$ to $q^{\prime}$ with shape $S$.\end{claim}

\subsection{A total order over the registers \label{sub:Noncomm:RegisterOrder}}

During the iteration step of the construction, we have to provide
function expressions for $\funcapptrad{\vector R_{S}^{\left(i+1\right)}}{q,q^{\prime}}$
in terms of the candidate function expressions at step $i$. Register
values may flow in complicated ways: consider for example the shape
in figure \ref{fig:Noncomm:Shapes:Weird}. The construction of $\funcapptrad{\vector R_{S}^{\left(i+1\right)}}{q,q^{\prime}}$
is greatly simplified if we assume that the shapes under consideration
are idempotent under concatenation.
\begin{defn}
\label{defn:Noncomm:RegisterOrder} Let $V$ be a finite set of registers,
and $\regleq$ be a total order over $V$. We call a shape $S$ over
$V$ \emph{normalized} with respect to $\regleq$ if
\begin{enumerate}
\item for all $u,v\in V$, if $v$ occurs in $\funcapptrad Su$, then $u\regleq v$,
\item for all $u,v\in V$, if $v$ occurs in $\funcapptrad Su$, then $u$
itself occurs in $\funcapptrad Su$, and
\item for all $v\in V$, there exists $u\in V$ such that $v$ occurs in
$\funcapptrad Su$.
\end{enumerate}
A $\ccrasimp$ $M$ is normalized if the shape of each of its update
expressions is normalized with respect to $\regleq$.
\end{defn}
For example, the shapes in figures \ref{fig:Noncomm:Shapes:q1a},
\ref{fig:Noncomm:Shapes:q1bb}, \ref{fig:Noncomm:Shapes:NoncommS1},
and \ref{fig:Noncomm:Shapes:NoncommS2} are normalized, while \ref{fig:Noncomm:Shapes:q1b}
and \ref{fig:Noncomm:Shapes:Weird} are not. Informally, the first
condition requires that all registers in the $\ccrasimp$ flow upward,
and the second ensures that shapes are idempotent. Observe that if
the individual transitions in a path are normalized, then the whole
path is itself normalized. 
\begin{prop}
\label{prop:Noncomm:RegisterOrder} For every $\ccrasimp$ $M$, there
is an equivalent normalized $\ccrasimp$ $M^{\prime}$.
\end{prop}
\begin{IEEEproof}
Let $M=\brtuple{Q,\Sigma,V,\delta,\mu,q_{0},F,\nu}$. Let $V^{\prime}=\rusetbr{x_{i}}{0\leq i\leq\left|V\right|}$
(so that $\left|V^{\prime}\right|=\left|V\right|+1$), and define
the register ordering as $x_{i}\regleq x_{j}$ iff $i\leq j$. $x_{0}$
is a sink register which accumulates all those register values which
are lost during computation. Let $Q^{\prime}$ be the set of all those
pairs $\brtuple{q,f}$, where $q\in Q$ is the current state, and
the permutation $\func fV{V^{\prime}\setminus\rosetbr{x_{0}}}$ is
the register renaming function. For simplicity, let us extend each
register renaming function $f$ to $\arrow{\union V{\Gamma}}{\union{V^{\prime}}{\Gamma}}$
by defining $\funcapptrad f{\gamma}=\gamma$, for $\gamma\in\Gamma$.
We further extend it to $\arrow{\kstar{\left(\union V{\Gamma}\right)}}{\kstar{\left(\union{V^{\prime}}{\Gamma}\right)}}$
by $\fabr f{v_{1}v_{2}\ldots v_{k}}=\funcapptrad f{v_{1}}\funcapptrad f{v_{2}}\ldots\funcapptrad f{v_{k}}$.
Let $F^{\prime}=\rusetbr{\brtuple{q,f}}{q\in F}$, and define the
output function $\nu^{\prime}$ as $\funcapptrad{\nu^{\prime}}{q,f}=\funcapptrad f{\funcapptrad{\nu}q}$.

For each state $\brtuple{q,f}\in Q^{\prime}$, and each symbol $a\in\Sigma$,
define $f^{\prime}$ as follows. For each register $v\in V$, if at
least one register occurs in $\fabr{\mu}{q,a,v}$, then $\funcapptrad{f^{\prime}}v=\min\rusetbr{\fabr fu}{u\mbox{ occurs in }\funcapptrad{\mu}{q,a,v}}$.
Observe that, because of the copyless restriction, for every pair
of distinct registers $u,v\in V$, $\funcapptrad{f^{\prime}}u\neq\funcapptrad{f^{\prime}}v$.
For all registers $v$ such that $\funcapptrad{f^{\prime}}v$ is still
undefined, define $\funcapptrad{f^{\prime}}v$ arbitrarily such that
$f^{\prime}$ is a permutation. Now $\fabr{\delta^{\prime}}{\brtuple{q,f},a}=\brtuple{\fabr{\delta}{q,a},f^{\prime}}$.

Define $\fabr{\mu^{\prime}}{\brtuple{q,f},a,x_{0}}=x_{0}+\funcapptrad f{v_{1}}+\funcapptrad f{v_{2}}+\cdots+\funcapptrad f{v_{k}}$,
where $\rosetbr{v_{1},v_{2},\ldots,v_{k}}$ is the set of registers
in $M$ whose value is lost during the transition. For all registers
$v\in V$, if $\fabr{\mu}{q,a,v}=v_{1}v_{2}\ldots v_{k}\in\kstar{\left(\union V{\Gamma}\right)}$,
define $\fabr{\mu^{\prime}}{\brtuple{q,f},a,\fabr{f^{\prime}}v}=\fabr f{v_{1}}+\fabr f{v_{2}}+\cdots+\fabr f{v_{k}}$.

For an arbitrary ordering $v_{1}\leq v_{2}\leq\cdots\leq v_{\left|V\right|}$
of the original registers $V$, define $\funcapptrad{f_{0}}{v_{i}}=x_{i}$.
It can be shown that the $\ccrasimp$ $M^{\prime}=\brtuple{Q^{\prime},\Sigma,V^{\prime},\delta^{\prime},\mu^{\prime},\brtuple{q_{0},f_{0}},F^{\prime},\nu^{\prime}}$
is equivalent to $M$, and that its transitions are normalized.\end{IEEEproof}

We will now assume that all $\ccrasimp$s and shapes under consideration
are normalized, and we elide this assumption in all definitions and
theorems.

\subsection{A partial order over shapes \label{sub:Noncomm:ShapeOrder}}

We now make the observation that some shapes cannot be used in the
construction of other shapes. Consider the shapes $S_{1}$ and $S_{\top}$
from figure \ref{fig:Noncomm:Shapes}. Let $\pi$ be a path through
the $\ccrasimp$ with shape $S_{1}$. Then, no sub-path of $\pi$
can have shape $S_{\top}$, because if such a sub-path were to exist,
then the value in register $y$ would be promoted to $x$, and the
registers $x$ and $y$ could then never be separated. We now create
a partial-order $\shleq$, and an equivalence relation $\suppeq$
over the set $\S$ of upward flowing shapes which together capture
this notion of ``can appear as a subpath''.
\begin{defn}
\label{defn:Noncomm:ShapeOrder} If $S$ is a shape over the set of
registers $V$, then the support of $S$, $\support S=\ruset{v\in V}{v\mbox{ occurs in }\funcapptrad Sv}$.
If $S_{1}$ and $S_{2}$ are two shapes, then $S_{1}\shlt S_{2}$
iff $\support{S_{1}}\supset\support{S_{2}}$. We call two shapes $S_{1}$
and $S_{2}$ \emph{support-equal}, written as $S_{1}\suppeq S_{2}$,
if $\support{S_{1}}=\support{S_{2}}$.
\end{defn}
For example, the shape $S_{\bot}$ from figure \ref{fig:Noncomm:Shapes}
is the bottom element of $\shleq$, and $S_{\top}$ is the top element.
$S_{1}\suppeq S_{2}$, and both shapes are strictly sandwiched between
$S_{\bot}$ and $S_{\top}$. Note that support-equality is a finer
relation than incomparability%
\footnote{Note that incomparability with respect to $\shleq$ is not even an
equivalence relation over shapes.%
} with respect to $\shleq$. In an early attempt to create a partial
order over shapes, we considered formalizing the relation $R_{\autobox{\mathit{sp}}}$,
``can appear as the shape of a sub-path''. However, this approach
fails because $R_{\autobox{\mathit{sp}}}$ is not a partial order.
In particular, observe that $\shapecat{S_{1}}{S_{2}}=S_{1}$, and
$\shapecat{S_{2}}{S_{1}}=S_{2}$. Thus, both $\tuple{S_{1},S_{2}}$
and $\tuple{S_{2},S_{1}}$ occur in $R_{\autobox{\mathit{sp}}}$,
and they are not equal. The presence of ``crossing edges'' in the
visualization of $S_{2}$ is what complicates the construction, but
we could not find a syntactic transformation on $\ccrasimp$s that
would eliminate these crossings.
\begin{claim}
\label{clm:Noncomm:SubpathLeq} Let $\pi$ be a path through the $\ccrasimp$
$M$ with shape $S$, and $\pi^{\prime}$ be a subpath of $\pi$ with
shape $S^{\prime}$. If $S^{\prime}\not\shlt S$, then $S^{\prime}\suppeq S$.
\end{claim}
\begin{IEEEproof}
Assume otherwise, so $S^{\prime}\not\suppeq S$. Then, for some register
$v\in\support S$, $v\notin S^{\prime}$. The effect of the entire
path $\pi$ is to make the initial value of $v$ flow into itself,
but on the subpath $\pi^{\prime}$, $v$ is promoted to some upper
register $v^{\prime}$. Because of the normalization condition (definition
\ref{defn:Noncomm:RegisterOrder}), it follows that on the suffix,
the value in $v^{\prime}\reglt v$ cannot flow back into $v$, leading
to a contradiction.\end{IEEEproof}

\begin{claim}
\label{clm:Noncomm:SubpathPrefix} Let $\pi$ be a path through the
$\ccrasimp$ $M$ with shape $S$, and let $\pi^{\prime}$ be the
shortest prefix with shape $S^{\prime}$ such that $S^{\prime}\not\shlt S$.
Then $S^{\prime}=S$.
\end{claim}
\begin{IEEEproof}
Assume otherwise. From claim \ref{clm:Noncomm:SubpathLeq}, we know
that $S^{\prime}\suppeq S$.
\begin{casenv}
\item For some register $u\notin\support S$, and registers $v,w\in\support S$,
with $v\neq w$, $u\to v$ in $S$, and $u\to w$ in $S^{\prime}$.
Once $u$ has flowed into $w$, the ``superpath'' cannot remove
$u$ from $w$. It is thus a contradiction that $u$ flows into $v$
in $S$.
\item For some register $v\in\support S$, the order of registers in $\funcapptrad Sv$
and $\funcapptrad{S^{\prime}}v$ are different. For some registers
$u$ and $w$, $u$ occurs before $w$ in $\funcapptrad Sv$, and
$w$ occurs before $u$ in $\funcapptrad{S^{\prime}}v$. However,
once the values of $w$ and $u$ have been appended to $v$ in the
order $wu$, they cannot be separated to be recast in the order $uw$.
It is thus a contradiction that $u$ occurs before $w$ in $\funcapptrad Sv$.\end{casenv}
\end{IEEEproof}

\subsection{Kleene-{*} and revisiting states \label{sub:Noncomm:Kstar}}

\global\long\def\Lfirst#1{\funcapptrad{L_{\autobox{\mathit{first}}}}{#1}}

\global\long\def\evfirst#1{\vector A_{#1}}

\global\long\def\evlast#1{\vector B_{#1}}

\global\long\def\evinc#1{\vector C_{#1}}

At each step of the iteration, for each pair of states $q,q^{\prime}\in Q$,
and for each shape $S$, we construct a new expression vector $\paregv{i+1}Sq{q^{\prime}}$,
summarizing paths in $\pareg{i+1}q{q^{\prime}}$ with shape $S$.
Recall that, from the DFA-to-regex translator, $\pareg{i+1}q{q^{\prime}}=\pareg iq{q^{\prime}}+\pareg iq{q_{i+1}}\kstar{\pareg i{q_{i+1}}{q_{i+1}}}\pareg i{q_{i+1}}{q^{\prime}}$.

Let $\evlast S$ be an expression vector which summarizes paths in
$\kstar{\pareg i{q_{i+1}}{q_{i+1}}}$ with shape $S$. We can then
write $\paregv{i+1}Sq{q^{\prime}}=\choice{\paregv iSq{q^{\prime}}}{\evinc S}$,
where $\evinc S=\mbox{ }\choiceop\rusetbr{\paregv i{S_{1}}q{q_{i+1}}\cdot\evlast{S_{2}}\cdot\paregv i{S_{3}}{q_{i+1}}{q^{\prime}}}{\shapecat{S_{1}}{\shapecat{S_{2}}{S_{3}}}=S}$.
Our goal is therefore to construct $\evlast S$, for each $S$. We
construct these expression vectors inductively, according to the partial
order $\shleq$. The remaining subsections are devoted to expressing
$\evlast S$.

\subsection{Decomposing loops \label{sub:Noncomm:Decomp}}

\begin{figure*}
\begin{centering}
\begin{tikzpicture}

  \node [state]                  (q1) {$q_{i + 1}$};
  \node [state, right=1.3 of q1] (q2) {$q_{i + 1}$};
  \node [state, right=1.3 of q2] (q3) {$q_{i + 1}$};
  \node [state, right=1.3 of q3] (q4) {$q_{i + 1}$};
  \node [state, right=1.3 of q4] (q5) {$q_{i + 1}$};
  \node [state, right=1.3 of q5] (q6) {$q_{i + 1}$};
  \node [state, right=1.3 of q6] (q7) {$q_{i + 1}$};
  \node [state, right=1.3 of q7] (q8) {$q_{i + 1}$};

  \path [->] (q1) edge node [label=above:{$S_1 = S$}, label=below:{$\sigma_1$}] {} (q2);
  \path [->] (q2) edge node [label=above:{$S_2 \suppeq S$}, label=below:{$\sigma_2$}] {} (q3);
  \path [->] (q3) edge node [label=above:{$\cdots$}] {} (q4);
  \path [->] (q4) edge node [label=above:{$S_j \suppeq S$}, label=below:{$\sigma_j$}] {} (q5);
  \path [->] (q5) edge node [label=above:{$\cdots$}] {} (q6);
  \path [->] (q6) edge node [label=above:{$S_k \suppeq S$}, label=below:{$\sigma_k$}] {} (q7);
  \path [->] (q7) edge node [label=above:{$S_f \shlt S$}, label=below:{$\sigma_f$}] {} (q8);

  \node [state, below=0.75 of q2] (qj0) {$q_{i + 1}$};
  \node [state, below=0.75 of q5] (qji) {$q_{i + 1}$};
  \node [state, below=0.75 of q7] (qjf) {$q_{i + 1}$};

  \path [-, dashed]  (q4)  edge (qj0);
  \path [-, dashed]  (q5)  edge (qjf);
  \path [->] (qj0)
             edge node [label=above:{$S_{\autobox{\mathit{pre}}} \shlt S$},
                        label=below:{$\sigma_{\autobox{\mathit{pre}}} \in \kstar{\pareg{i}{q_{i + 1}}{q_{i + 1}}}$}] {}
             (qji);
  \path [->] (qji)
             edge node [label=above:{$S_{\autobox{\mathit{suff}}}$},
                        label=below:{$\sigma_{\autobox{\mathit{suff}}} \in \pareg{i}{q_{i + 1}}{q_{i + 1}}$}] {}
             (qjf);

\end{tikzpicture}
\par\end{centering}

\caption{\label{fig:Noncomm:Decomp} Decomposing paths in $\kstar{\pareg i{q_{i+1}}{q_{i+1}}}$
with shape $S$. $\sigma_{j}$ can be unambiguously written as $\sigma_{\autobox{\mathit{pre}}}\sigma_{\autobox{\mathit{suff}}}$,
with $\sigma_{\autobox{\mathit{suff}}}\in\pareg i{q_{i+1}}{q_{i+1}}$.}
\end{figure*}
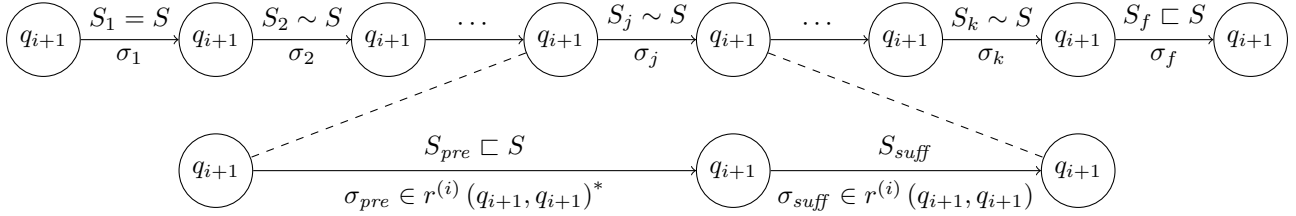

Consider any path $\sigma$ in $\kstar{\pareg i{q_{i+1}}{q_{i+1}}}$
with shape $S$. From claims \ref{clm:Noncomm:SubpathLeq} and \ref{clm:Noncomm:SubpathPrefix},
we can unambiguously decompose $\sigma=\sigma_{1}\sigma_{2}\ldots\sigma_{k}\sigma_{f}$,
where
\begin{enumerate}
\item each $\sigma_{j}\in\kstar{\pareg i{q_{i+1}}{q_{i+1}}}$ is a self-loop
at $q_{i+1}$,
\item for each $j$, $1\leq j\leq k$, the shape $S_{j}$ of $\sigma_{j}$
is support-equal to $S$, $S_{j}\suppeq S$, and $S_{f}\shlt S$,
and
\item for each $j$, $1\leq j\leq k$, and for each proper prefix $\sigma_{\autobox{\mathit{pre}}}\in\kstar{\pareg i{q_{i+1}}{q_{i+1}}}$
of $\sigma_{j}$, $S_{\autobox{\mathit{pre}}}\shlt S$.
\end{enumerate}
Let us call the split $\sigma=\sigma_{1}\sigma_{2}\ldots\sigma_{k}\sigma_{f}$
the \emph{$S$-decomposition} of $\sigma$. See figure \ref{fig:Noncomm:Decomp}.

Consider some shape $S^{\prime}\suppeq S$, and let $\Lfirst{S^{\prime}}$
be the set of all paths $\pi\in\kstar{\pareg i{q_{i+1}}{q_{i+1}}}$
with shape $S^{\prime}$ such that no proper prefix $\pi_{\autobox{\mathit{pre}}}$
of $\pi$ has shape $S_{\autobox{\mathit{pre}}}\suppeq S$. We can
then unambiguously write $\pi=\pi_{\autobox{\mathit{pre}}}\pi_{\autobox{\mathit{last}}}$,
with $\pi_{\autobox{\mathit{pre}}}\in\kstar{\pareg i{q_{i+1}}{q_{i+1}}}$,
$\pi_{\autobox{\mathit{last}}}\in\pareg i{q_{i+1}}{q_{i+1}}$, and
such that $S_{\autobox{\mathit{pre}}}\shlt S$. Define $\evfirst{S^{\prime}}=\mbox{ }\choiceop\rusetbr{\paregv{i+1}{S_{\autobox{\mathit{pre}}}}{q_{i+1}}{q_{i+1}}\cdot\paregv i{S_{\autobox{\mathit{post}}}}{q_{i+1}}{q_{i+1}}}{\shapecat{S_{\autobox{\mathit{pre}}}}{S_{\autobox{\mathit{post}}}}=S\mbox{ and }S_{\autobox{\mathit{pre}}}\shlt S}$.
\begin{claim}
\label{clm:Noncomm:Kstar1:Lfirst} For all shapes $S^{\prime}\suppeq S$,
the expression vector $\evfirst{S^{\prime}}$ summarizes all paths
in $\Lfirst{S^{\prime}}$.
\end{claim}

\subsection{Computing $\evlast S$ \label{sub:Noncomm:BS}}

We now construct the expression vector $\evlast S$. Consider a path
$\sigma$, and its $S$-decomposition $\sigma=\sigma_{1}\sigma_{2}\ldots\sigma_{k}\sigma_{f}$.
Given a register $v$, and a patch $1\leq k\leq\left|\funcapptrad Sv\right|+1$,
three cases may arise:

First, if $\funcapptrad Sv=\strempty$, i.e. $v$ is reset during
the computation. $v$ was reset while processing $\sigma_{k}$. Any
registers flowing into it during this time were also reset by $\sigma_{k}$.
Thus, its value is entirely determined entirely by $\sigma_{k}$ and
$\sigma_{f}$. First define $\vector F=\mbox{ }\choiceop\rusetbr{\evfirst{S_{1}}\cdot\evlast{S_{2}}}{\shapecat{S_{1}}{S_{2}}=S\mbox{ and }S_{2}\shlt S}$,
and let $L_{f}=\bigunion{S^{\prime}\suppeq S}{\Lfirst{S^{\prime}}}$.
Observe that $\funcapptrad{\mu}{q_{i+1},\sigma,v}=\funcapptrad{\mu}{q_{i+1},\sigma_{k}\sigma_{f},v}=\funcapptrad{\vector F_{v,1}}{\sigma_{k}\sigma_{f}}$,
and therefore define $\evlast{S,v,1}=\splitsum{\const{\kstar{L_{f}}}{\strempty}}{\vector F_{v,1}}$.

Second, if $1<k<\left|\funcapptrad Sv\right|+1$, i.e. that $k$ refers
to an internal patch in $\funcapptrad Sv$. Once the registers are
combined in some order, any changes can only be appends at the beginning
and end of the register value. The $k^{\mbox{th}}$ constant in $\funcapptrad{\mu}{q_{i+1},\sigma,v}$
is consequently determined by $\sigma_{1}$. Therefore, define $\evlast{S,v,k}=\splitsum{\evfirst{S,v,k}}{\const{\kstar{L_{f}}}{\strempty}}$.

\begin{figure}
\begin{centering}
\begin{tikzpicture}

  \node [state]                  (q1) {$q_{i + 1}$};
  \node [state, right=1.8 of q1] (q2) {$q_{i + 1}$};
  \node [state, right=1.8 of q2] (q3) {$q_{i + 1}$};

  \node [below=0.35 of q1] (x1) {$v$};
  \node [below=0.35 of q2] (x2) {$v$};
  \node [below=0.35 of q3] (x3) {$v$};

  \node [below=0.35 of x1] (y1) {$w$};
  \node [below=0.35 of x2] (y2) {$w$};

  \path [->] (q1) edge node [label=above:{$S_j \suppeq S$}, label=below:{$\sigma_j$}] {} (q2);
  \path [->] (q2) edge node [label=above:{$S_{j + 1} \suppeq S$}, label=below:{$\sigma_{j + 1}$}] {} (q3);

  \path [->] (x1) edge (x2);
  \path [->] (x2) edge (x3);

  \path [(->] (y1) edge (y2);
  \path [->] (y2) edge (x3);

  \path [-, dashed] (q1) edge (x1);
  \path [-, dashed] (x1) edge (y1);
  \path [-, dashed] (q2) edge (x2);
  \path [-, dashed] (x2) edge (y2);
  \path [-, dashed] (q3) edge (x3);

\end{tikzpicture}
\par\end{centering}

\caption{\label{fig:Noncomm:BS} For any path in $\kstar{\pareg i{q_{i+1}}{q_{i+1}}}$,
inward flows into a (support) register $v$ have to be from non-support
registers.}
\end{figure}
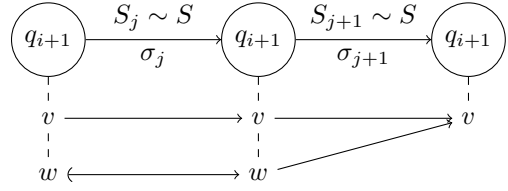

Finally, if $k=1$, or $k=\left|\funcapptrad Sv\right|+1$, i.e. $k$
is either the first or the last patch. First, we know that $v\in\support S$.
Also, we know that any registers which flow into $v$ have to be non-support
registers. See figure \ref{fig:Noncomm:BS}. Thus, the value being
appended to $v$ while processing $\sigma_{j}$ is determined entirely
by $\sigma_{j}$ and $\sigma_{j-1}$. The idea is to use chained sum
to compute this value. 

We will now define $\evlast{S,v,k}$ for $k=\left|\funcapptrad Sv\right|+1$.
The case for $k=1$ is symmetric, and would involve reversing the
order of the operators, and replacing chained sum with the left-chained
sum.

We are determining the constant value appended to the end of $v$
while processing $\sigma$. We distinguish three phases of addition:
while processing $\sigma_{1}$, only the constant at the end of $\evfirst{S,v,k}$
is appended. While processing $\sigma_{j}$, $j>1$, \emph{both string
constants and registers} appearing after the occurrence of $v$ in
$\funcapptrad Sv$ are appended. Third, while processing $\sigma_{f}$,
both string constants and registers appearing after the occurrence
of $v$ in $\funcapptrad Sv$ are appended. The interesting part about
the second case is that this appending happens in a loop, and we therefore
need the lookback provided by the chained sum operator. Otherwise,
this case is similar to the simpler third case, where a value is appended
exactly once.

While processing $\sigma_{1}$, some symbols are appended to the $k^{\mbox{th}}$
position in $\funcapptrad Sv$. This is given by $f_{\autobox{\mathit{pre}}}=\splitsum{\evfirst{S,v,k}}{\const{\kstar{L_{f}}}{\strempty}}$.

Similarly, while processing the suffix $\sigma_{f}$, some symbols
are appended. Say some register $u\to v$ in $S_{f}$. Then $u\notin\support{S_{f}}$,
and hence $u\notin\support S$ and $u\notin\support{S_{k}}$. Thus,
the value appended by $\sigma_{f}$ is determined by $\sigma_{k}\sigma_{f}$.
For each pair of shapes $S_{k}$ and $S_{f}$ such that $S_{k}\suppeq S$,
and $S_{f}\shlt S$, consider $\evfirst{S_{k}}^{\prime}=\lshift{\evfirst{S_{k}}}{\domain{S_{f}}}$,
and $\evlast{S_{f}}^{\prime}=\rshift{\evlast{S_{f}}}{\domain{\evfirst{S_{k}}}}$.
Consider the update expression $\evlast{S_{f},v}^{\prime}$: say this
is $v:=\sigma v\tau$, where $\sigma$ and $\tau$ are strings over
expressions and registers. For each register $u$ in $\tau$, substitute
the value $\evfirst{S_{k},u,1}^{\prime}$ -- since $u$ was reset
while processing $S_{k}$, this expression gives the contents of the
register $u$ -- and interpret string concatenation in $\tau$ as
the function combinator sum. Label this result as $f_{\autobox{\mathit{post}},S_{k},S_{f}}$.
Define $f_{\autobox{\mathit{post}}}=\splitsum{\const{\kstar{L_{f}}}{\strempty}}{\choiceop\rusetbr{f_{\autobox{\mathit{pre}},S_{k},S_{f}}}{S_{k}\suppeq S\mbox{ and }S_{f}\shlt S}}$.

Finally, consider the value appended while processing $\sigma_{j}$,
for $j>1$. This is similar to the case for $\sigma_{f}$: if $u\to v$
in $S_{j}$, when $u\notin\support{S_{j}}$ and $u\notin\support{S_{j-1}}$.
Thus, the value appended by $\sigma_{j}$ is determined by $\sigma_{j-1}\sigma_{j}$.
For each pair of states $S_{j-1}\suppeq S$ and $S_{j}\suppeq S$,
consider $\evfirst{S_{j-1}}^{\prime}=\lshift{\evfirst{S_{j-1}}}{\domain{\evfirst{S_{j}}}}$,
and $\evfirst{S_{j}}^{\prime}=\rshift{\evfirst{S_{j}}}{\domain{\evfirst{S_{j-1}}}}$.
Consider the update expression $\evfirst{S_{j},v,k}^{\prime}$. Let
this be $v:=\sigma v\tau$, where $\sigma$ and $\tau$ are strings
over expressions and registers. For each register $u$ in $\tau$,
substitute the value $\evfirst{S_{j-1},u,1}^{\prime}$ -- since $u$
was reset while processing $S_{j-1}$, this expression gives the contents
of the register $u$ -- and interpret string concatenation in $\tau$
as the function combinator sum. Label this result as $f_{S_{j-1},S_{j}}$.
Define $f=\sum(\choiceop\rusetbr{f_{S_{j-1},S_{j}}}{S_{j-1}\suppeq S\mbox{ and }S_{j}\suppeq S},L_{f})$.

Finally, define $\evlast{S,v,k}=\choice{\p{\repsum{f_{\autobox{\mathit{pre}}}}{f_{\autobox{\mathit{post}}}}}}{\p{\repsum{f_{\autobox{\mathit{pre}}}}{\repsum f{f_{\autobox{\mathit{post}}}}}}}$.

By construction, we have:
\begin{claim}
$\evlast S$ summarizes all paths in $\kstar{\pareg i{q_{i+1}}{q_{i+1}}}$
with shape $S$.
\end{claim}
This completes the proof of theorem \ref{thm:Noncomm}.

\section{Conclusion \label{sec:Conclusion}}

In this paper, we have characterized the class of regular functions
that map strings to values from a monoid using a set of function combinators.
We hope that these results provide additional evidence of robust and
foundational nature of this class. The identification of the combinator
of chained sum, and its role in the proof of expressive completeness
of the combinators, should be of particular technical interest. There
are many avenues for future research. First, the question whether
all the combinators we have used are \emph{necessary} for capturing
all regular functions remains open (we conjecture that the set of
combinators is indeed minimal). Second, it is an open problem to develop
the notion of a congruence and a Myhill-Nerode-style characterization
for regular functions (see \cite{Boj13} for an attempt where authors
give such a characterization, but succeed only after retaining the
``origin'' information that associates each output symbol with a
specific input position). Third, it would be worthwhile to find analogous
algebraic characterizations of regularity when the domain is, instead
of finite strings, infinite strings \cite{AFT12} or trees \cite{EM99,AD12}
and/or when the range is a semiring \cite{DKV09,CRA-LICS}. Finally,
on the practical side, we plan to develop a declarative language for
document processing based on the regular combinators identified in
this paper.

\bibliographystyle{plain}
\bibliography{references}

\end{document}